\documentclass[10pt]{article}
\usepackage{opex3}
\usepackage{amsmath}
\usepackage{amssymb}
\usepackage{epsfig}
\newcommand{\nm}{ \,\text{nm}}
\newcommand{\f}{ f }
\newcommand{\mum}{ \,\mu\text{m}}
\newcommand{\bra}[1]{\langle #1 | }
\newcommand{\ket}[1]{|  #1 \rangle }
\newcommand{\hTE}{\text{hybrid-TE}}
\newcommand{\hTM}{\text{hybrid-TM}}
\newcommand{\amode}{\text{I}}
\newcommand{\bmode}{\text{II}}
\newcommand{\Rneff}{n_i^{-1}}

\begin{document}
\title{Efficient spectroscopy of single embedded emitters using optical fiber taper waveguides}

\author{Marcelo Davan\c co$^{1,2}$ and Kartik Srinivasan$^1$}
\address{$^1$Center for Nanoscale Science and Technology,
National Institute of Standards and Technology, Gaithersburg, MD, 20899-6203\\
$^2$Maryland NanoCenter,\\ University of Maryland, College Park, MD,
20742} \email{mdavanco@nist.gov}

\begin{abstract}
A technique based on using optical fiber taper waveguides for
probing single emitters embedded in thin dielectric membranes is
assessed through numerical simulations. For an appropriate membrane
geometry, photoluminescence collection efficiencies in excess of
10$~\%$ are predicted, exceeding the efficiency of standard
free-space collection by an order of magnitude. Our results indicate
that these fiber taper waveguides offer excellent prospects for
performing efficient spectroscopy of single emitters embedded in
thin films, such as a single self-assembled quantum dot in a
semiconductor membrane.
\end{abstract}

\bibliographystyle{osa}



\section{Introduction}

The development of novel techniques for efficient detection and
spectroscopy of individual quantum emitters in the solid state is
essential for an understanding of the emitter and its relationship
with the surrounding environment. Challenges involved with
spectroscopy of such systems include tight focusing requirements and
separation of the signal of interest from background scattered light
due to the host crystal. In recent years, tools such as high
numerical aperture objectives
\cite{gerardot_APL,vamivakas.nano.letters.7.2892,Wrigge.Nature.Phys.4.60}
and near-field scanning optical microscopy tips
\cite{gerhardt.prl.033601} have been used for both efficient
photoluminescence (PL) collection and resonant optical spectroscopy.

A related problem is optical spectroscopy of microphotonic
resonators, where wavelength - scale focusing and effective
separation of the signal from background scattered light are also
necessary. The optical fiber taper waveguide is a tool that has been
investigated for such experiments. These structures, sometimes
referred to as silica nanofibers, are standard single mode optical
fibers that have been heated and stretched down to a
wavelength-scale minimum diameter. At such small dimension, the
evanescent field of the waveguide mode extends into the air cladding
and can be used to interrogate surrounding structures. By conducting
the tapering process adiabatically and symmetrically, a double-ended
device with standard single mode fiber input and output can be
produced, with a typical overall transmission loss of less than
10$~\%$.  The combination of low loss, single mode guidance, and
access to a wavelength-scale evanescent field has made tapered fiber
probes invaluable for a wide number of microcavity-based
experiments, including those in single atom and quantum dot cavity
quantum electrodynamics
\cite{aoki.nat,srinivasan.nat,srinivasan.pra.033839}. These
properties also suggest that they can be effective tools for single
emitter spectroscopy. Indeed, theoretical work
\cite{LeKien.pra.72.032509,klimov.PhysRevA.69.013812} has indicated
that silica nanofibers can be used to efficiently collect
fluorescence from single gas phase atoms, and experimental progress
to this end has been made \cite{Nayak_OE}.

The above studies indicate that optical fiber taper waveguides can
be a general spectroscopic tool for nanophotonic systems, providing
motivation for the present work. Here, we study their potential for
efficient non-resonant and resonant fluorescence collection
experiments on single solid state emitters embedded in a dielectric
slab. Additional motivation is provided in previous experimental
work \cite{srinivasan:091102}, where a fiber taper waveguide was
used in non-resonant PL measurements of a single InAs quantum dot
embedded in a 256 nm thick GaAs membrane on top of an
Al$_{0.7}$Ga$_{0.3}$As pedestal. There, a luminescence collection
efficiency of $\approx0.1~\%$ was estimated based on the measured
saturated photon count rates, neglecting possible radiation rate
modification due to the host structure, a non-resonant microdisk
cavity, and any quantum dot non-idealities. Important questions left
unanswered include how close this efficiency was to the theoretical
maximum for the particular geometry considered, and how the geometry
may be modified to lead to higher efficiencies. Furthermore, the
possibility of studying resonant fluorescence, not addressed in~
\cite{srinivasan:091102}, should also be considered.  As described
in other recent works
\cite{gerardot_APL,vamivakas.nano.letters.7.2892,
Wrigge.Nature.Phys.4.60,gerhardt.prl.033601,srinivasan.pra.033839,Muller.prl},
resonant spectroscopy has advantages in comparison to non-resonant
PL measurements in terms of greatly improved spectral resolution,
potentially improved temporal resolution, the ability to avoid
generation of decoherence-inducing excess carriers, well-defined
state preparation, and possible utility in quantum information
processing experiments.

\begin{figure}[t]
\centerline{\includegraphics[scale=0.6]{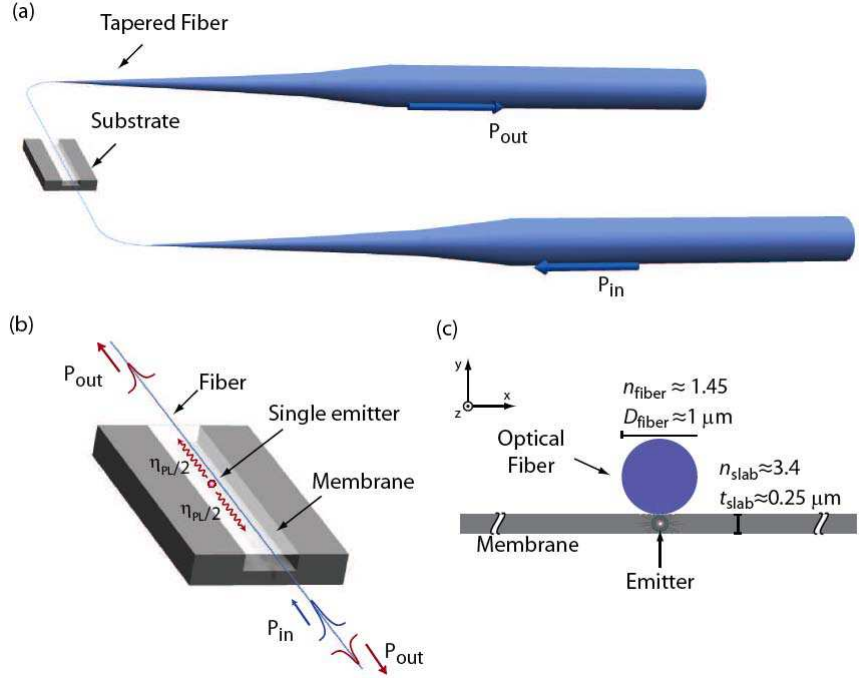}} \caption{(a)
Single emitter probing setup based on a tapered
  fiber waveguide. (b) Detail of (a), showing the membrane that carries the
  emitter.  (c) Schematic of the substrate cross-section, showing
  membrane and fiber.} \label{fig-sexy-scheme}
\end{figure}

To address these points, we consider the spontaneous emission of a
dipole embedded in a semiconductor slab surrounded by air and in
contact with a micron diameter silica optical fiber. Specifically,
we envision an experimental measurement setup as depicted in Fig.
\ref{fig-sexy-scheme}, where the optical fiber taper waveguide is
brought into close proximity with a thin semiconductor membrane
hosting a single quantum emitter.  The fiber has been adiabatically
tapered to reach the single-mode condition over its central region
(hundreds of microns long). The emitter may be optically pumped
through an external fiber-coupled source, and radiates in both
directions into the fiber, with a total coupling efficiency
$\eta_{\text{PL}}$ into the fundamental fiber mode (both backward
and forward directions). We perform full wave, 3D finite difference
time domain (FDTD) electromagnetic simulations of a single classical
dipole radiating in the probing structure shown in
Figs.~\ref{fig-sexy-scheme}(b)-\ref{fig-sexy-scheme}(c), and obtain
the fiber-collected power. We find that fiber collection
efficiencies in excess of $\approx 12~\% (30~\%)$ may be achieved
for dipole moments horizontally (vertically) oriented with respect
to the membrane plane, exceeding what may be achieved with standard
free-space optics PL collection schemes by an order of magnitude or
more. These results suggest that optical fiber taper waveguides are
potentially quite valuable in future studies of single solid state
quantum emitters.

To determine ways of further improving collection efficiency, we
analyze the dependence of both the modified spontaneous emission
rate and PL collection efficiency on the electromagnetic supermodes
of the probing structure.  The starting point of our analysis is
~\cite{LeKien.pra.72.032509}, which determines the modified
spontaneous emission rate of a multi-level atom near a silica fiber.
Addition of the semiconductor slab is a non-trivial modification of
this problem, requiring the use of finite-element method (FEM)
simulations to compute the guided and radiation modes of the
composite slab-fiber system. These modes are used along with FDTD to
determine the individual supermode contributions to the spontaneous
emission fiber coupling fraction $\eta_{\text{PL}}$, under the
assumption that the embedded emitter can be approximated as a
two-level system.

The paper is organized as follows. In Section
\ref{section_fiber_probing}, we use FDTD to estimate the spontaneous
emission collection efficiency into an optical fiber for a single
quantum dot embedded in a semiconductor membrane, as shown in Fig.
 \ref{fig-sexy-scheme}. The results of this section are analyzed in
Section \ref{section_analysis} in terms of the propagating
supermodes of the probing structure. In Section
\ref{section_discussion}, we further discuss the physical
interpretation of these results and consider how they change with
fiber size and if the host semiconductor membrane rests on a
substrate (i.e., a non-undercut slab geometry). We also consider a
pair of specific configurations with improved collection
efficiencies, for which additional investigations are underway.
Finally, we describe how resonant fluorescence measurements can be
performed through our fiber-based probing and collection scheme.

\section{Fiber-based embedded single emitter photoluminescence collection}
\label{section_fiber_probing} 
\begin{figure}[t]
\centerline{\includegraphics[scale=0.65,trim = 0 10 0
20]{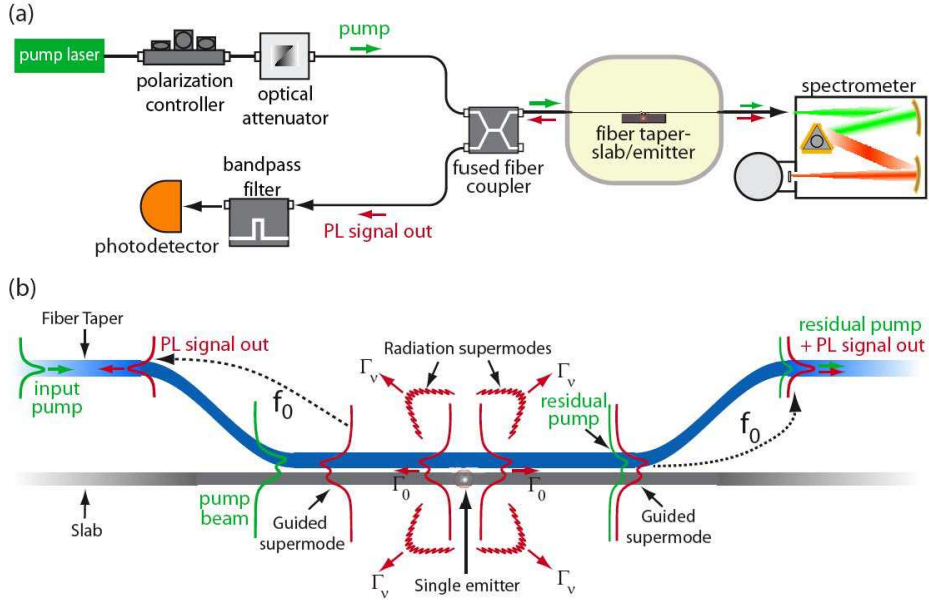}} \caption{(a) Envisioned experimental configuration
for fiber-based non-resonant photoluminescence (PL) spectroscopy
with a tapered fiber waveguide. Emitted light is coupled into both
forward and backward channels of the fiber taper waveguide, and can
be wavelength resolved with a grating spectrometer or spectral
filter. (b) Schematic of single emitter excitation and PL collection
via the tapered fiber probe. A non-resonant pump signal is injected
into the input fiber and converted into a guided supermode of the
composite waveguide, illuminating the slab-embedded dipole. The
dipole radiates into guided and radiative supermodes, with rates
  $\Gamma_0$ and $\Gamma_\nu$, respectively. Power is transferred with efficiency $\f_0$ from the
  supermode to the fiber mode and vice-versa.}  \label{fig-coupler}
\end{figure}

In this section, we analyze the
photoluminescence (PL) collection from a single emitter embedded in
a dielectric slab with the fiber taper probe setup illustrated in
Fig. \ref{fig-sexy-scheme}.  For non-resonant PL measurements,
application of the fiber probe in the configuration of Fig.
\ref{fig-coupler}(a) is envisioned, which provides fiber-confined
paths for excitation and PL collection in both forward and backward
directions. As indicated in Fig. \ref{fig-sexy-scheme}(b), the
micron-scale single mode region of the optical fiber waveguide is
brought into close proximity with the top surface of a dielectric
membrane that hosts the emitter, over a length of several to
hundreds of wavelengths. Fiber and membrane together form the
composite dielectric waveguide with cross-section shown in Fig.
\ref{fig-sexy-scheme}(c). The composite waveguide supports a
complete set of guided, leaky and radiation supermodes
\cite{snyder.love.waveguides}, originating from the hybridization of
fiber and slab modes. As depicted in Fig. \ref{fig-coupler}(b), part
of the non-resonant pump power initially carried by the fiber is
coupled to a guided supermode of the composite waveguide and
transferred toward the emitter. Illuminated by the pump light, the
emitter radiates (at a red-shifted wavelength) into modes of the
composite fiber-slab waveguide.  Our objective is to determine the
fraction of this emission that would be measured at the single mode
fiber output, after the micron-scale fiber has transitioned away
from the slab.  The portion of the total emission collected by the
fiber is carried by those supermodes with sufficient transverse
confinement to reach the fiber transition regions with significant
amplitude. Thus, apart from guided modes, leaky modes may contribute
significantly to the total PL collection.

Within this scenario, we estimate the spontaneous emission
collection efficiency for a single emitter in a geometry that
approximates the tapered-fiber-based quantum dot spectroscopy setup
of ~\cite{srinivasan:091102}. In that work, PL spectroscopy of
single self-assembled InAs quantum dots embedded in a non-resonant
256 nm thick GaAs microdisk was demonstrated, with an estimated
collection efficiency of $0.1~\%$ (including both channels of the
optical fiber). Our numerical estimate here is obtained through full
electromagnetic simulations of a point dipole radiating inside the
composite waveguide of Fig. \ref{fig-sexy-scheme}(c).  The geometry
consists of a dielectric slab of index $n_{\text{slab}}=3.406$, and
fiber of radius $r=500\nm$ and refractive index
$n_{{\text{fiber}}}=1.45$ placed directly on the top surface of the
slab. The dipole is assumed to radiate at $\lambda=1.3\mum$, a
wavelength at which the fiber possesses, apart from the fundamental
guided mode, a near-cutoff mode which is not significant in our
calculations. Slab membranes of thicknesses varying between $100\nm$
and $260\nm$ were investigated, sufficient to avoid quantum dot
dephasing due to surface-state-related processes, as suggested in
\cite{wang.APL.3423}.  In Section \ref{section_discussion}, we
consider how these results change when $r$ is decreased to 300 nm
and when the slab membrane is placed directly on a substrate.

\subsection{Simulation Method}
 \label{section_sim_methods}
To estimate the PL collection efficiency of our fiber-based
probing scheme, we considered a single classical electric dipole
radiating in the composite dielectric waveguide of Fig.
\ref{fig-sexy-scheme}(c). An upper bound for the percentage of the
total emitted power $P_{\text{Tot.}}$ coupled to the fundamental
optical fiber mode at an arbitrary position $z$ along the guide is
\begin{equation}
\eta_{\text{PL}} = 2\frac{P_z}{P_{\text{Tot.}}}f_{\text{fiber}},
\label{eq_eta_PL_fdtd}
\end{equation}
where $P_z$ is the power flowing normally through the constant-$z$
plane, and $f_{\text{fiber}}$ is the overlap integral~\cite{ref:Huang3}
\begin{equation}
\f_{\text{fiber}}=\frac{ \text{Re}\left\{
\iint_{S}(\mathbf{e}_f\times\mathbf{h}^*)\cdot\hat{z}\,dS \,
\iint_{S}(\mathbf{e}\times\mathbf{h}_f^*)\cdot\hat{z}\,dS\right\} }
{ \text{Re}\left\{
\iint_{S}(\mathbf{e}_f\times\mathbf{h}_f^*)\cdot\hat{z}\,dS \right\}
\, \text{Re}\left\{
\iint_{S}(\mathbf{e}\times\mathbf{h}^*)\cdot\hat{z}\,dS \right\} }.
\label{eq_overlap}
\end{equation}
Here, $\left\{{\mathbf{e}},{\mathbf{h}}\right\}$ are the
steady-state radiated fields at position $z$ and
$\left\{{\mathbf{e}_f},{\mathbf{h}_f}\right\}$ the fundamental fiber
mode fields. The factor of 2 in Eq.~(\ref{eq_eta_PL_fdtd}) is
introduced to account for collection from both fiber ends.

The steady-state fields were obtained through FDTD simulations
\cite{lumerical}. The symmetry of the geometry allowed us to choose
either anti-symmetric ($\hat{\mathbf{x}}\times\mathbf{E}=0$) or
symmetric ($\hat{\mathbf{x}}\times\mathbf{H}=0$) boundary conditions
on the $yz$-plane, the former being used for $x$-polarized dipoles,
and the latter for $y$- and $z$-polarized dipoles. The computational
domain was cubic with a 10 $\mum$ side and perfectly-matched layers
(PMLs) were used around the domain limits to simulate an open
domain. Dipole excitation consisted of a $\lambda=1.3\mum$ carrier
modulated with a $0.1\nm$ bandwidth gaussian envelope. Simulations
ran until no field amplitude could be detected in the domain. The
total radiated power $P_{\text{Tot.}}$ was calculated by adding the
integrated steady-state power through all the computational window
sides. The PL collection efficiency $\eta_{\text{PL}}$ was
calculated at several $z$-planes along the propagation direction
($z$) in the computational domain.

\subsection{Simulation Results }
\subsubsection{Total radiated power}
\label{section-total-power}
\begin{figure}[h]
\centerline{\includegraphics[scale=0.65, trim = 0 10 0
20]{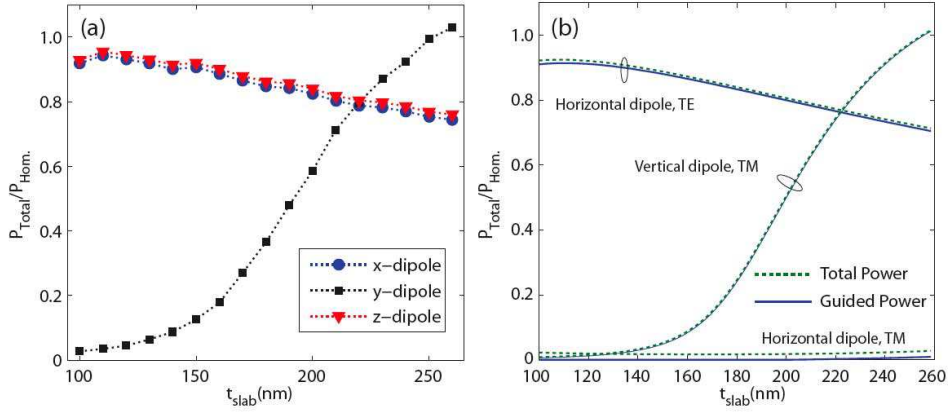}} \caption{(a) Total radiated power at $\lambda =
1.3 \mum$ for a dipole in the composite slab-fiber waveguide,
normalized to the radiated power in a homogeneous medium of index
$n_{\text{slab}}$. (b) Guided and total radiated powers into TE (TM)
waves for a horizontal (vertical and horizontal) dipole at the
center of a dielectric slab with $n_{\text{slab}}=3.406$. }
\label{fig_total_rates}
\end{figure}

Figure \ref{fig_total_rates}(a) shows
the FDTD-calculated, total radiated power ($P_{\text{Tot.}}$) for
$x$-, $y$- and $z$-oriented dipoles located at the slab center in
the structure of Fig. \ref{fig-sexy-scheme}(c). The curves were
normalized to the total radiated power of an electric dipole current
source $\mathbf{J}(\mathbf{r},t)=-i\omega\mathbf{p} \delta
(\mathbf{r}) e^{-i\omega_0t}$ in a homogeneous dielectric medium of
refractive index $n_{\text{slab}}$ and electric dipole moment
$\mathbf{p}$ \cite{Jackson.dipole.power}:
\begin{equation}
P_{\text{Hom.}}=\frac{\mu_0}{4\pi}n_{\text{slab}}|\mathbf{p}|^2\frac{\omega_0^4}{3c}.
\label{eq_hom_radiated_power}
\end{equation}

\begin{figure}[h]
\centerline{\includegraphics[scale=0.7, trim = 0 10 0
20]{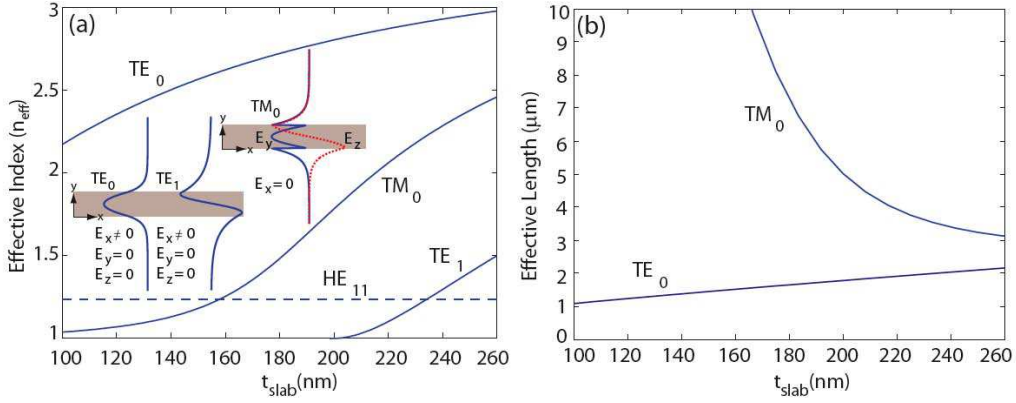}} \caption{(a) Effective indices of the slab TE$_0$,
TE$_1$ and TM$_0$ modes at $\lambda = 1.3 \mum$. The HE$_{11}$ fiber
mode effective index is also shown. Inset: field components of
TE$_0$, TE$_1$ and TM$_0$ slab modes. (b) Effective lengths,
$L_{\text{eff}}$, for the TE$_0$ and TM$_0$ slab modes. }
\label{fig_neff}
\end{figure}

The total radiated power is not significantly different when the
fiber is absent. This is evident in Fig. \ref{fig_total_rates}(b),
where the power radiated by vertical or horizontal dipoles into
modes of an isolated slab are plotted. These curves were calculated
using the transfer-matrix method of ~\cite{Benisty.JOSAA.98}.
Transverse-electric waves (TE) have electric field components in the
slab plane only, while transverse-magnetic (TM) modes have only
in-plane magnetic field components.  As a result, a vertical
($y$-directed) dipole excites only TM waves, while horizontal ($x$-
and $z$-directed) dipoles can excite both TE and TM waves
\cite{Benisty.JOSAA.98}, although as we see in Fig.
\ref{fig_total_rates}(b), the TE mode excitation is much more
significant. The similarity between the total radiated power curves
for the slab with or without the fiber indicates its weak
perturbative effect on the emission, so we first proceed by
analyzing the isolated slab.  Figure \ref{fig_neff}(a) displays the
effective indices of the TE$_0$, TE$_1$ and TM$_0$ modes, for
varying slab thickness. The evolution of the field concentration at
the dipole for both TE and TM modes may be inferred from the
effective length,
$L_{\text{eff}}=\int\epsilon|\mathbf{E}|^2\,dy/|\mathbf{E(\mathbf{r}_0)}|$,
where $\mathbf{r}_0$ is the dipole position, plotted in Fig.
\ref{fig_neff}(b). The effective length is inversely proportional to
the dipole power coupled to the mode
\cite{Rigneault.PhysRevA.54.2356,Urbach.PhysRevA.57.3913}.

Given the large refractive index of the slab relative to air, total
internal reflection leads to preferential dipole excitation of
guided waves. This is verified in the curves for guided and total
radiated powers in Fig. \ref{fig_total_rates}(b)
\cite{Benisty.JOSAA.98}. For a horizontal dipole at the slab center,
most of the radiated power is carried by the TE$_0$ mode, which has
a relatively small effective length, $L_{\text{eff}}$. The TE$_1$
mode is not excited, since its electric field has a node at the
dipole location. The TM$_0$ mode is also not excited, as
$\mathbf{E}_x=0$ and $\mathbf{E}_z$ has a null at the slab center.
For a vertical dipole, excitation of the TM$_0$ mode is more
effective, especially since no TE modes are generated
\cite{Benisty.JOSAA.98}. The strong emission inhibition for vertical
dipoles at small $t_{\text{slab}}$ is due to poor outcoupling into
the air, resulting from total internal reflection, and poor field
concentration, apparent in the large TM$_0$ effective lengths
$L_{\text{eff}}$ of Fig. \ref{fig_neff}(b). As a result, even if
high fiber collection efficiencies can be achieved, this strong
suppression of spontaneous emission can lead to low overall photon
count rates at the detector.

\subsubsection{Photoluminescence Collection}

\begin{figure}[!b]
\centerline{\includegraphics[scale=0.7,trim = 0 10 0
20]{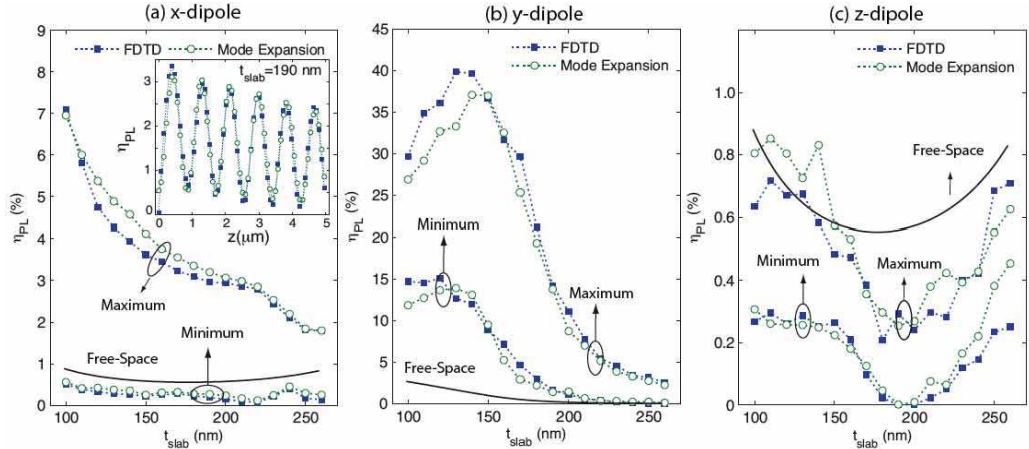}} \caption{Fiber-collected PL coupling
efficiency $\eta_{\text{PL}}$ for (a) $x$-, (b) $y$- and (c) $z$-
oriented dipoles at $\lambda=1.3 \mum$. Collection through both the
forward and backward fiber channels is considered (notice changing
axis scales). Since $\eta_{\text{PL}}$ oscillates with $z$, its
maximum and minimum along $z<5\mum$ are plotted. The inset in (a)
shows $\eta_\text{PL}$ as function of $z$. Squares: FDTD results.
Circles: Mode-expansion results. Solid lines: Free-space collection
efficiency considering a NA=0.7 objective.} \label{fig_efficiency}
\end{figure}

Using Eq.~(\ref{eq_eta_PL_fdtd}), the collection efficiency
$\eta_{\text{PL}}$ was calculated for a range of propagation lengths
a few wavelengths away from the dipole location ($1\mum<z<5\mum$).
The results reveal strong periodic oscillations as a function of
this distance, as shown in the inset of
Fig.~\ref{fig_efficiency}(a). As we shall discuss in the following
section, this oscillation is due to the beating of two principal
waveguide supermodes with distinct propagation constants,
corresponding to the sloshing of power between fiber and slab, as
commonly observed in coupled waveguide systems \cite{ref:Huang3}.
Oscillations in $\eta_{\text{PL}}$ are accompanied by a decrease in
its peak value - as we shall see in Section \ref{section_analysis},
this decay is due to leaky modes with propagations lengths of
several wavelengths. The maximum and minimum values of
$\eta_{\text{PL}}$ within the simulated propagation range were taken
to generate the PL efficiency plots for $x$-, $y$- and $z$-directed
dipoles of Fig.~\ref{fig_efficiency}. Note that the values plotted
in Fig.~\ref{fig_efficiency} include collection through both the
forward and backward channels of the fiber, an experimentally viable
situation through use of a fused fiber coupler
(Fig.~\ref{fig-coupler}(a)).

We first consider the results for in-plane oriented dipoles (in the
$x$ and $z$ directions), which are of relevance to quantum dot
experiments such as in ~\cite{srinivasan:091102}.  For $x$-oriented
dipoles, the maximum collection efficiency ranges from $\approx7~\%$
to $\approx2~\%$ for $t_{\text{slab}}$ varying from $100\nm$ to
$260\nm$, while the minimum efficiencies remain close to 0.1$~\%$
throughout. For $z$-oriented dipoles, the efficiency varies below
1$~\%$ for all thicknesses.  In ~\cite{srinivasan:091102}, a single
quantum dot inside a 256-$\nm$ GaAs slab was probed, and a measured
collection efficiency of 0.1$~\%$ was estimated. This value is
within the limits predicted by Figs.~\ref{fig_efficiency}(a)
and~\ref{fig_efficiency}(c), given that the in-plane dipole
orientation was not known, and the quantum dot was assumed to have
unity radiative efficiency. In addition, the in-plane position of
the quantum dot was such that the underlying Al$_{0.7}$Ga$_{0.3}$As
support pedestal could have had significant influence on the PL
collection.  We consider its effects in Section
\ref{section_discussion}.

To further gauge our current results, we use the method of
~\cite{Benisty.JOSAA.98} to plot, in Fig.~\ref{fig_efficiency}, the
maximum achievable free-space collection efficiency for a dipole
embedded in the slab (without the fiber), using a $0.7$ numerical
aperture (NA) objective.  We see that the maximum achievable
$\eta_{\text{PL}}$ for horizontally-oriented dipoles in the
fiber-based collection scheme is superior to the free-space
collection efficiency by as much as an order of magnitude, as
indicated in Fig.~\ref{fig_efficiency}(a). As is the case with other
co-propagating waveguide directional couplers \cite{ref:Huang3},
control of the interaction length is necessary to achieve the most
efficient power transfer. Here, to obtain the highest values of
$\eta_{\text{PL}}$, one might envision controlling the length of the
membrane through standard planar fabrication techniques.
Alternately, the effective fiber-membrane interaction region can be
limited using 'dimpled' fiber taper waveguides \cite{michael.OE}.
Given the low propagation losses in the optical fiber and a typical
insertion loss of $10~\%$ through the taper, the present technique
constitutes a competitive option for PL spectroscopy.

For vertically oriented dipoles, the collection efficiency may reach
above $35~\%$ at $t_{\text{slab}}=140\nm$, where the dipole emission
is strongly suppressed. For thicker slabs, the suppression is
reduced, but the collection efficiency drops, reaching $\approx
3~\%$ for $t_{\text{slab}}=250\nm$. Even this collection efficiency
is nonetheless an order of magnitude higher than what may be
obtained with free-space collection with a $0.7$-NA objective,
plotted in Fig. \ref{fig_efficiency}(b). Although self-assembled
quantum dots in semiconductor slabs are typically in-plane oriented
dipoles, we note that manipulation of the quantum dot growth can
change its shape sufficiently to produce structures that support
TM-polarized emission \cite{Jayavel}. The results presented here
would be of significant interest to such systems.

While the results presented in this section are analyzed in detail
in Section~\ref{section_analysis}, we provide here a brief
explanation for the relatively high collection efficiencies found,
regarding the fiber-slab system as a directional coupler. In a
directional coupler, power transfer takes place between waves, in
respective waveguides, of similar propagation constants. Although
transfer is most efficient when both waves are perfectly matched,
small but significant transfer may still occur between mismatched
modes. As discussed in Section~\ref{section-total-power}, most of
the x-polarized dipole radiation is carried by TE slab mode waves. A
significant portion of these waves, though phase-mismatched to the
fiber mode, transfer small amounts of power to the latter,
collectively leading to relatively high (8 \% maximum) collection
efficiencies. The substantially higher (40 \% maximum) collection
efficiency obtained for y-oriented dipoles results from a more
effective power transfer from TM slab waves to the fiber mode: phase
matching between the fiber and TM$_0$ slab modes is apparent in
Fig.~\ref{fig_neff}(a), for slab thicknesses close to 160nm.

\section{Analysis}
\label{section_analysis}

While the FDTD simulations provide estimates of the total
spontaneous emission rate and fiber-coupled collection efficiency,
they do not necessarily provide physical understanding of effects
such as the oscillation and decay in PL collection efficiency as a function of
separation from the dipole in the propagation direction, or the
insight needed to extend this work to modified geometries.  In this
section, we address these issues through FEM simulations that
determine the modal structure of the composite fiber-slab system.
This allows us to determine the spontaneous emission rate, and contribution to
the total collected PL,
of individual supermodes of the system. This approach furthermore allows us
to estimate the evolution of the coupling efficiency for coupling lengths much
longer than practical in FDTD simulations.

\subsection{Supermodes of the composite waveguide}
The composite waveguide of Fig. \ref{fig-sexy-scheme} supports a set
of hybrid supermodes of the fundamental, micron-scale fiber mode and
the bound, TE$_m$ ($E_y=0$) or TM$_m$ ($H_y=0$) slab modes. These
supermodes are henceforth referred to as $\hTE_m$ or $\hTM_m$. The
fiber creates a region of increased effective refractive index on
the slab plane, spanning a discrete set of laterally confined $\hTE$
or $\hTM$ supermodes. A continuum of radiation $\hTE$ or $\hTM$
supermodes also exists, which represents waves not laterally
confined by the increased fiber-induced effective index. A portion
of this radiative supermode spectrum corresponds to optical power
that, although unconfined, lingers near the fiber for relatively
long propagation distances (tens of wavelengths), and may contribute
significantly to the PL collection efficiency. These wave bundles
may be represented by a discrete set of leaky supermodes, which
extend infinitely in the lateral direction and decay in the
propagation direction \cite{snyder.love.waveguides}. The relevant
waves for PL collection within our scheme are thus the guided and
leaky $\hTE$ or $\hTM$ supermodes.

The supermode field distributions of our structure were obtained
through the vector eigenvalue problem for the magnetic field
$\mathbf{H}$
\begin{equation}
\nabla\times\left( \frac{1}{\epsilon(\mathbf{r})}
\nabla\times\mathbf{H} \right)-\left(\frac{\omega}{c}\right)^2=0.
\label{eq_mode_eigen}
\end{equation}
Here, $\mathbf{H}=\mathbf{H}(\mathbf{r})\exp(\xi z)$, and $\xi =
i\beta-\alpha_z = i\omega n_\text{eff}/c $ is a complex propagation
constant with a phase term $\beta$ and a decay term $\alpha_z$, or
equivalently a complex effective index $n_\text{eff}$. The number of
free-space wavelengths necessary for the supermode amplitude to
decay by a factor~$0<\delta<1$ is $
N_{\delta}=-\Rneff\ln(\delta)/2\pi$, where $\Rneff =
\text{Im}\{n_\text{eff}\}^{-1}$. This number is used in our analysis
to roughly determine the probing length (i.e., the length over which
the fiber contacts the slab) for supermode power collection to be
most effective. Since the dipole is aligned with the center of the
fiber, the problem is symmetric with respect to the $yz$ plane. An
$x$-oriented dipole thus only excites $\hTE$ supermodes, for which
$\hat{\mathbf{x}}\times\mathbf{E}=0$ on the symmetry plane.
Conversely, $y$- and $z$-oriented dipoles only excite $\hTM$
supermodes, for which $\hat{\mathbf{x}}\times\mathbf{H}=0$ on the
$yz$ plane.

The isolated slab waveguide supports a TE$_0$ and a TM$_0$ mode for
all slab thicknesses $t_{\text{slab}}$ considered, and a TE$_1$ mode
for $t_{\text{slab}}\gtrapprox200\nm$ (Fig. \ref{fig_neff}), so that
the composite waveguide structure supports $\hTE_0$ and $\hTM_0$ and
$\hTE_1$ modes in the same ranges. Field profiles for the
strongest-confinement $\hTE_0$ and $\hTM_0$ supermodes are shown in
Figs.~\ref{fig_mode_profiles}(a) and~\ref{fig_mode_profiles}(d) of
Section \ref{section_sim_results}.

\subsection{Modified spontaneous emission rate in the composite waveguide}
To describe the spontaneous emission of a single quantum emitter
into the composite dielectric waveguide of Fig.
\ref{fig-sexy-scheme} in terms of its supermodes, we begin by making
use of ~\cite{LeKien.pra.72.032509}, where the spontaneous emission
rate of a multi-level atom into the guided and radiative modes of a
silica optical fiber are calculated. These expressions are derived
from the Heisenberg equations for a multilevel atom experiencing a
dipole-type interaction with a vacuum field reservoir that is
described in terms of propagating waveguide modes.
While~\cite{LeKien.pra.72.032509} focused exclusively on optical
fibers, here we deal with general dielectric waveguides, so that the
expressions we use are modified accordingly from their originals,
with the derivation presented in Appendix B.

The waveguide, extending in the $z$ direction, is open, and thus
supports a finite set of guided modes and a continuum of
unconfined, radiation modes \cite{snyder.love.waveguides}. Guided
modes and radiation modes are respectively labeled with the
indices $\mu$ and $\nu$, and have propagation constants
$\beta_\mu$ and $\beta$ at frequency $\omega$. For the radiation
mode continuum, $\beta$ may assume any value in the interval
$|\beta|<\omega n/c$, where $n$ is the refractive index of the
medium surrounding the waveguide. Evanescent radiation modes are
excluded, as they do not participate in the radiation process
\cite{sondegaard.PhysRevA.64.033812}.

For a two-level atom located in a general dielectric waveguide at
a position $\mathbf{r_0}$, with a transition of energy
$\hbar\omega_0$ and dipole moment $\mathbf{p}$, the spontaneous
emission rates into guided and radiative modes are, respectively,
$\Gamma_\mu = 2\pi|G_\mu(\omega_0)|^2$ and $\Gamma_\nu =
2\pi|G_\nu(\omega_0)|^2$, where
\begin{eqnarray}
G_\mu = \sqrt{\frac{\omega\beta_\mu'}{4\pi\epsilon_0\hbar}}\left
[\mathbf{p}\cdot\mathbf{e}_\mu(\mathbf{r_0})\right]\label{eq_G_mu_body}
\qquad\text{and}\qquad G_\nu = \sqrt{\frac{\omega}{4\pi
N_\nu\hbar}}\left
[\mathbf{p}\cdot\mathbf{e}_\nu(\mathbf{r_0})\right].\label{eq_G_nu_body}
\end{eqnarray}
Here, $\mathbf{e}_\mu$ and $\mathbf{e}_\nu$ are the electric fields
of the guided and radiative modes, with respective propagation
constants $\beta_\mu$ and $\beta$.  Also, $\beta'_\mu =
\partial\beta_\mu/\partial\omega$ and $N_\nu$ is such that \cite{snyder.love.waveguides}
\begin{equation}
\iint\limits_S dS\,\left(
\mathbf{e}_\nu\times\mathbf{h}_{\nu'}^*\right)\cdot\hat{\mathbf{z}}=N_\nu\delta(\beta-\beta')
.
\end{equation}
Guided and radiative modes are normalized so that
\begin{equation}
\iint\limits_S dS\,\epsilon(\mathbf{r})|\mathbf{e}_\mu|^2=1
\label{eq_guided_mode_norm}
\end{equation}
\begin{equation}
 \iint\limits_S
dS\epsilon(\mathbf{r})(\mathbf{e}_\nu\cdot\mathbf{e}_{\nu'}^*)_{\beta
=\beta',n=n'}=\delta(\omega-\omega') \label{eq_radiative_mode_norm}
\end{equation}
where $S$ is the cross-section of the waveguide and $\epsilon(\mathbf{r})$
the spatial permittivity distribution. The total emission rate is
\begin{eqnarray}
\Gamma = \sum_{\mu}\Gamma_\mu+\sum_{n}\int
d\beta\,\Gamma_\nu.\label{eq_Gamma}
\end{eqnarray}

According to Eq.~(\ref{eq_Sz_waveguide}), the power emitted from a
classical dipole into a particular mode depends on the modal
electric field concentration at the dipole position (the modal
electric fields are normalized to the power density flux integrated
over the waveguide cross-section). We introduce the effective area
$A_{\text{eff}}= \int\epsilon|\mathbf{E}|^2dS
/|\mathbf{E}(\mathbf{r}_0)|^2$ as a measure of the field
concentration at the emitter position. This expression is implicit
in the modal rates of Eq.~(\ref{eq_G_mu_body}), given the
normalization expressions, Eqs. (\ref{eq_guided_mode_norm}) and
(\ref{eq_radiative_mode_norm}). The total spontaneous emission rate,
being the sum of the individual modal rates, deviates considerably
from the free-space rate if highly concentrated modes (i.e., small
effective areas) exist at the emitter location.

To avoid calculation of all radiative modes of the composite
waveguide, the spontaneous emission rate $\Gamma$ was obtained
through the following relationship between the spontaneous emission
rates for an emitter in a homogeneous medium
($\Gamma_{\text{Hom.}}$) and into a waveguide
($\Gamma_{\text{WG}}$), and the total classical dipole radiated
powers in bulk ($P_{\text{Hom.}}$) and in a waveguide
($P_{\text{WG}}$) \cite{snyder.love.waveguides}:
\begin{equation}
\frac{\Gamma_{\text{WG}}}{\Gamma_{\text{Hom.}}}=\frac{P_{\text{WG}}}{P_{\text{Hom.}}}
\label{eq_spont_rate_frac}
\end{equation}
The validity of this expression is demonstrated in Appendix A (a
similar relation for cavities has been established in \cite{Xu:99}).
This equivalence allowed us to use FDTD simulations to determine
total spontaneous emission rates.

\subsection{Photoluminescence Collection}

With the help of Eq.~(\ref{eq_spont_rate_frac}), we may describe the
electric and magnetic fields radiated by a dipole located at $z=0$
inside the slab in terms of traveling supermodes. At an arbitrary
$z$-plane along the waveguide direction,
$\mathbf{e}=\sum_{m=1}^{M}\sqrt{\Gamma_m/\Gamma}\,\mathbf{e}_{m}\exp(i\xi_mz)$
and
$\mathbf{h}=\sum_{m=1}^{M}\sqrt{\Gamma_m/\Gamma}\,\mathbf{h}_{m}\exp(i\xi_mz)$,
where $\mathbf{e}_{m}$ and $\mathbf{h}_{m}$ are the fields of the
$m$-th eigensolution of Eq.~(\ref{eq_mode_eigen}), with eigenvalue
$\xi_m=\beta_m+i\alpha_{z,m}$. The factor $\Gamma_m$ is the
supermode emission rate from Eq.~(\ref{eq_G_mu_body}). Applied in
Eq.~(\ref{eq_overlap}), these fields lead to the following
expression for the percentage of the total spontaneous emission
carried by the fiber mode:

\begin{equation}
\eta_{PL} = 2 \sum_{m=1}^{M}f_m\frac{\Gamma_m}{\Gamma}+ 2
\sum_{\substack{m, n=1 \\ m\ne n}
}^{M}\sqrt{\frac{\Gamma_m}{\Gamma}}
\sqrt{\frac{\Gamma_n}{\Gamma}}\text{Re} \left\{\sqrt{f_m^h
f_n^e}\exp[i(\xi_m-\xi_n^*)z] \right\}, \label{eq_eta_PL_multimode}
\end{equation}
where
\begin{equation}
\sqrt{\f_{m}^h}=\frac{\iint_{S}(\mathbf{e}_f\times\mathbf{h}_m^*)\cdot\hat{z}\,dS}{\sqrt{\text{Re}
\left\{\iint_{S}(\mathbf{e}_m\times\mathbf{h}_m^*)\cdot\hat{z}\,dS\right\}\text{Re}
\left\{\iint_{S}(\mathbf{e}_f\times\mathbf{h}_f^*)\cdot\hat{z}\,dS\right\}}},
\label{eq_multimode_overlap_h}
\end{equation}

\begin{equation}
\sqrt{\f_{m}^e}=\frac{\iint_{S}(\mathbf{e}_m\times\mathbf{h}_f^*)\cdot\hat{z}\,dS}{\sqrt{\text{Re}
\left\{\iint_{S}(\mathbf{e}_m\times\mathbf{h}_m^*)\cdot\hat{z}\,dS\right\}\text{Re}
\left\{\iint_{S}(\mathbf{e}_f\times\mathbf{h}_f^*)\cdot\hat{z}\,dS\right\}}},
\label{eq_multimode_overlap_e}
\end{equation}


\noindent and $f_m=\text{Re}
\left\{\sqrt{\f_m^h}\sqrt{\f_m^e}\right\}$.  The factor of 2 is
introduced to account for collection from both fiber ends. While the
first sum in Eq.~(\ref{eq_eta_PL_multimode}) corresponds to the
individual supermode contributions to $\eta_{PL}$, the latter
corresponds to beating of the individual supermode field components.
These contributions, which may be substantial, are periodic in $z$,
with beat lengths $L_{\text{z}}=2\pi/(\beta_{{m}}-\beta_{{n}})$,
where $\beta_{{m,n}}$ are the $m$-th and $n$-th supermode
propagation constants. The $m$-th supermode individual contribution
to the spontaneous emission collection efficiency is
$\eta_{\text{PL},m}=\f_m\cdot(\Gamma_m/\Gamma)=\f_m\gamma_m$. The
ratio $\gamma_{\text{m}}=\Gamma_m/\Gamma$ gives the fraction of
spontaneous emission coupled into the supermode. Reflecting the fact
that supermodes in our structure are hybrids of the fundamental
fiber mode and slab TE and TM modes, $\f_m$ is called the fiber-mode
fraction of the $m$-th supermode.

The estimate of $\eta_{\text{PL},m}$ could be improved through a
more detailed consideration of the transition between fiber-slab
supermode and fundamental fiber mode.  In the case of an abrupt
transition (e.g., if the membrane is terminated at cleaved facets),
modal reflection and transmission coefficients may be obtained
through rigorous wave matching at the interface, yielding the power
transferred to the isolated fiber mode~\cite{ref:Huang3}. On the
other hand, a power transfer in excess of 90~\% is estimated from
the Fresnel reflection and transmission coefficients at such an
interface, considering the supermode and fiber effective indices.
Transmission levels of this magnitude were confirmed in FDTD
simulations of a truncated slab of thickness $190\nm$, for dipoles
placed at several distances from the transition. This suggests that
the mode overlap method used here and in the previously presented
FDTD results yields reasonable upper-bound estimates for the
achievable collection efficiency.

\subsection{Supermode Calculation Method}

The eigenvalue equation Eq.~(\ref{eq_mode_eigen}) was solved with
FEM with quadratic vector elements for $\mathbf{H}_x$ and
$\mathbf{H}_y$ and Lagrange elements for $\mathbf{H}_z$
\cite{comsol}. Given the symmetry of the problem, only half of the
geometry was considered, with either anti-symmetric
($\hat{\mathbf{x}}\times\mathbf{E}=0$) or symmetric
($\hat{\mathbf{x}}\times\mathbf{H}=0$) boundary conditions on the
$y$-axis, which automatically selected either $\hTE$ or $\hTM$
solutions. Perfectly-matched layers were defined over the outer
boundaries to simulate an open domain, as leaky modes were expected.
Perfect electric conductor ($\hat{\mathbf{n}}\times\mathbf{E}$=0)
boundary conditions surround the PML on the outermost boundaries.
The eigenvalue calculation produced a finite, discrete set of
supermode solutions, which were used in
Eq.~(\ref{eq_eta_PL_multimode}) to estimate $\eta_\text{PL}$. The
respective emission rates $\Gamma_m$ were calculated with the
guided-mode expression in Eq.~(\ref{eq_G_mu_body}). In all our
calculations, fifty supermodes were used, which was sufficient to
closely reproduce the FDTD results.

\subsection{Simulation results}
\label{section_sim_results}

\begin{figure}[t]
\centerline{\includegraphics[scale=0.7,trim = 0 10 0
20]{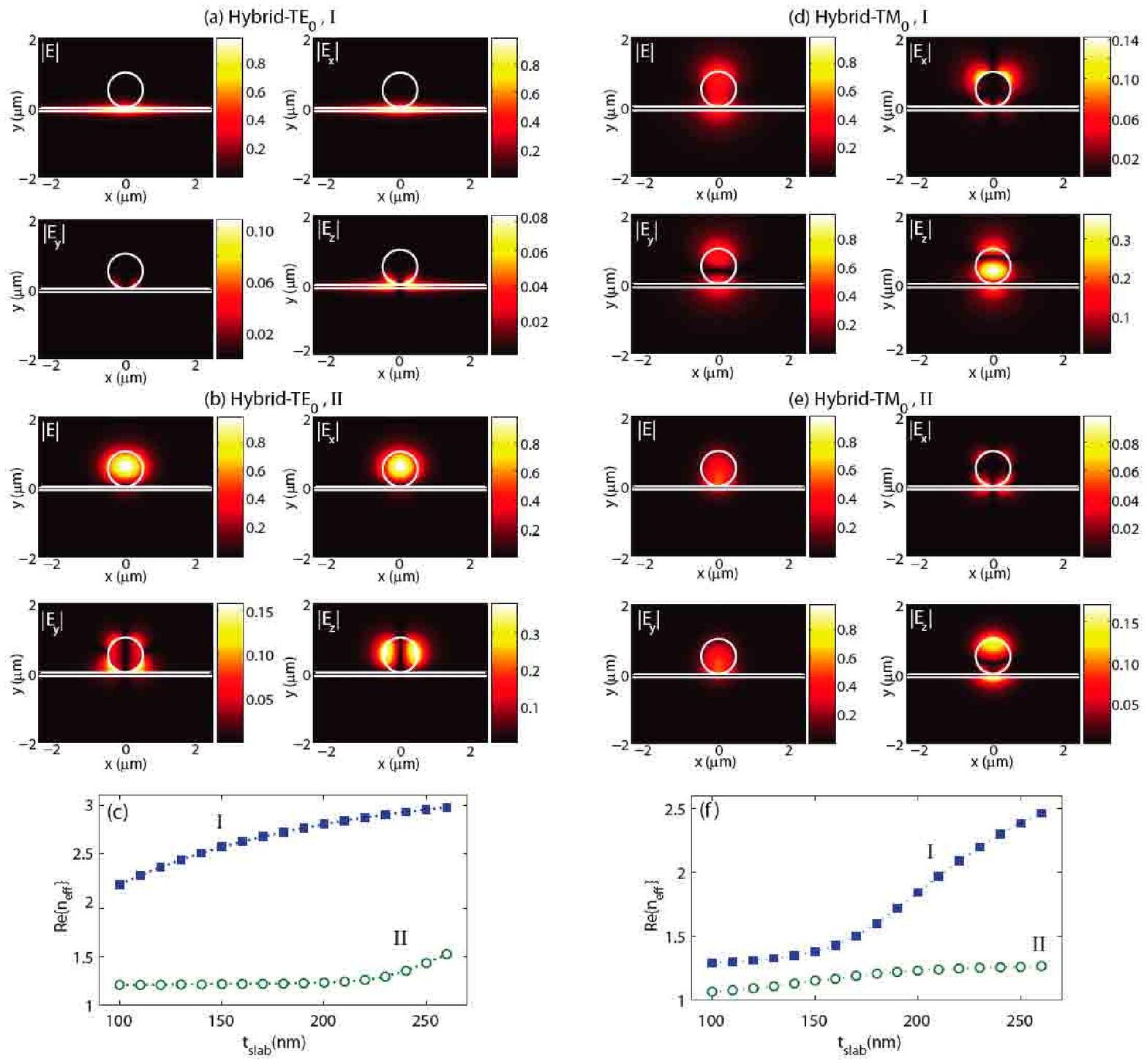}} \caption{Electric field distribution for
(a), (b): $\hTE_0$ types-$\amode$ (slab-like) and $\bmode$
(fiber-like) supermodes; (d), (e): $\hTM_0$ types-$\amode$  and
$\bmode$ supermodes. Fields are for $t_{\text{slab}}=100\nm$ and
$\lambda=1.3\mum$. (c) and (f): effective indices for (c) $\hTE_0$
and (f) $\hTM_0$ supermodes of types $\amode$ and $\bmode$ for
varying $t_\text{slab}$. Fields have been normalized to the total
field amplitude maxima in each case. Color scales are in arbitrary
units.} \label{fig_mode_profiles}
\end{figure}

The PL collection efficiencies for $x$-, $y$- and $z$-dipoles
calculated with the supermode expansion method above are plotted in
Fig. \ref{fig_efficiency}. The discrepancies in the $y$- and $z$
dipole cases relative to the FDTD results are largely due to an
insufficiently fine grid used in the FDTD simulations, where a full
3D simulation is run. In the FEM simulations, a cross-section of the
structure is analyzed and hence can be discretized with a finer
resolution. As discussed below, the main $\hTM$ modes, excited by
either $y$- or $z$-dipoles, are strongly concentrated in the region
between fiber and slab, requiring a very fine grid for high
accuracy. To avoid prohibitively long computation times, a coarser
grid was used.

\subsubsection*{Hybrid TE modes: $x$-oriented dipoles}
\begin{figure}[!t]
\centerline{\includegraphics[scale=0.7,trim = 30 10 0
10]{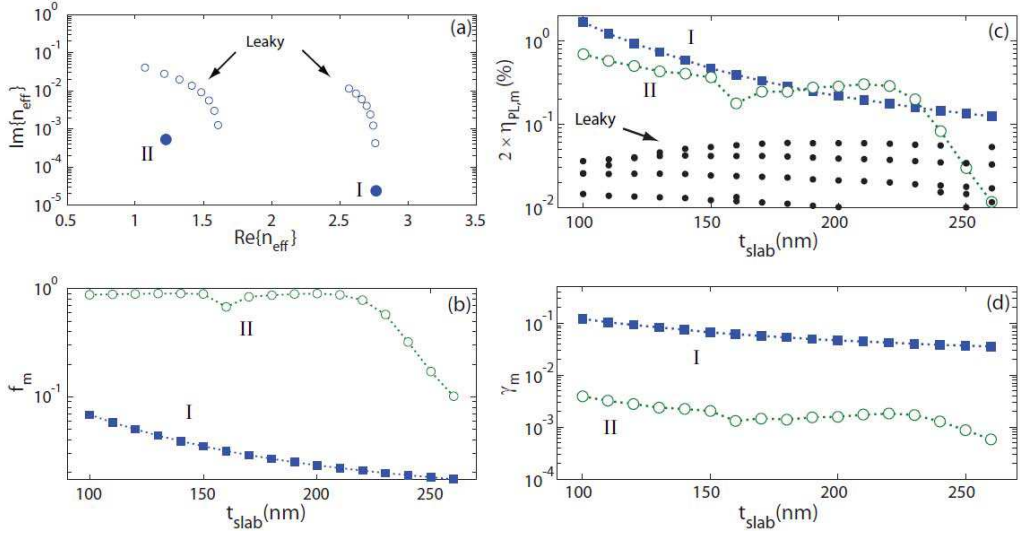}} \caption{ (a) Effective indices for $\hTE$
supermodes at $t_{\text{slab}}=190\nm$ and $\lambda=1.3 \mum$. The
two supermodes with highest modal emission rates are indicated with
filled circles labeled $\amode$ and $\bmode$, corresponding to
slab-like and fiber-like modes. Open circles indicate leaky modes.
(b) Fiber fractions for supermodes $\amode$ and $\bmode$. (c) PL
collection contributions from the individual supermodes, for an
$x$-oriented dipole moment. The factor of 2 is to account for
collection from both fiber ends. Black dots are for leaky modes.(d)
Modal emission ratios for supermodes $\amode$ and $\bmode$. Filled
squares/empty circles are for type $\amode$/$\bmode$ supermodes.}
\label{fig_gamma-asy}
\end{figure}

Figure \ref{fig_gamma-asy}(a) shows a representative effective index
distribution for $\hTE$ supermodes. For $x$-oriented dipole
excitation, the two supermodes labeled $\amode$ and $\bmode$ have
the largest contribution to the total collected PL. Types-$\amode$
and $\bmode$ supermodes are mostly concentrated in the slab and
fiber regions respectively, and are thus referred to as slab-like
and fiber-like. Representative electric-field profiles for these
main supermodes are shown in Figs.~\ref{fig_mode_profiles}(a) and
~\ref{fig_mode_profiles}(b), while Fig.~\ref{fig_mode_profiles}(c)
shows $\text{Re}\{n_\text{eff}\}$ as a function of $t_\text{slab}$
for these modes. The periodic oscillaton of $\eta_{\text{PL}}$ with
$z$, observed in the FDTD simulations, is very well reproduced by
the mode expansion calculations, as shown in the inset of
Fig.~\ref{fig_efficiency}(a). This oscillation is due to the beating
of the two main supermodes, and persists when leaky eigenmodes are
excluded from the calculation. The beat length is given by
$L_{\text{z}}=2\pi/(\beta_{\text{I}}-\beta_{\text{II}})$, where
$\beta_{\text{I,II}}$ are the respective propagation constants.

The supermode power contributions, $\eta_{\text{PL},m}$, are shown
in Fig.~\ref{fig_gamma-asy}(c). The slab-like supermode is dominant
for $t_{\text{slab}} \lessapprox 160\nm$ and $t_{\text{slab}}
\gtrapprox 240\nm$, being surpassed by the fiber-like mode at
intermediate thicknesses. In this range, the fiber-like supermode's
significantly larger fiber fraction (Fig.~\ref{fig_gamma-asy}(b)) is
enough to compensate for its much lower modal emission ratio
$\gamma_{0,\bmode}$ (Fig.~ \ref{fig_gamma-asy}(d)). Individually,
leaky supermodes pointed out in Fig.~\ref{fig_gamma-asy}(a) add
little to the total collected power, however their aggregate
influence is non-negligible: excluding leaky modes, the maximum
collected power drops between 40$~\%$ and 70$~\%$, a significant
portion of the total collected PL.

Among the leaky modes with $\eta_\text{PL,m}$ above 1$~\%$ of the
slab-like supermode contribution, the one with the lowest individual
contribution has $\Rneff\lessapprox156$, corresponding to
$N_{\delta=0.9}\gtrapprox2.6$ wavelengths for 90$~\%$ amplitude
decay. As a result, the highest efficiency collection shown in
Fig.~\ref{fig_efficiency}(a) will be achieved for coupling lengths
of a few microns; the contribution from leaky modes will begin to
drop out as the coupling lengths increase beyond this.  The slab-
and fiber-like supermodes have $N_{\delta=0.9}\gtrapprox427$ and
$N_{\delta=0.9}\gtrapprox22$, respectively, so that a coupling
length of $\lessapprox30\mum$ continues to allow collection of PL
coupled to both main supermodes. Finally, collection that is at
least as high as the slab-like supermode's contribution in
Fig.~\ref{fig_gamma-asy}(a) is achievable for lengths shorter than
550 $\mum$.  We note that this latter result is still on par with
what can be achieved through free-space collection
(Fig.~\ref{fig_efficiency}(a)).

\subsubsection*{Hybrid TM modes: $y$- and $z$-oriented dipoles}

\begin{figure}[!t]
\centerline{\includegraphics[scale=0.7,trim = 30 10 0
10]{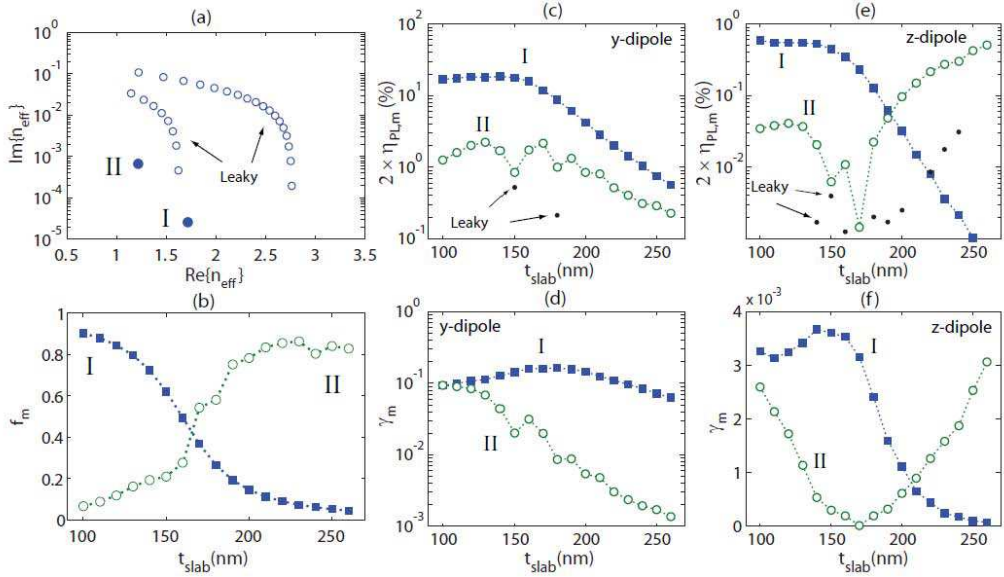}} \caption{ (a) Effective indices for $\hTM$
supermodes at $t_{\text{slab}}=190\nm$ and $\lambda=1.3 \mum$. The
two supermodes with highest modal emission rates are indicated with
filled circles labeled $\amode$ and $\bmode$. Open circles indicate
leaky modes. (b) Fiber fractions for supermodes $\amode$ and
$\bmode$. (c) and (e): PL collection contributions from individual
supermodes, for $y$- and $z$-oriented dipole moments respectively
(notice changing axis scales). The factor of 2 is to account for
collection from both fiber ends. (d) and (f): Modal emission ratios,
for $y$- and $z$-oriented dipole moments of type-$\amode$ and
type-$\bmode$ supermodes  (notice changing axis scales). Filled
squares/empty circles are for type $\amode$/$\bmode$ supermodes;
black dots are for leaky modes. } \label{fig_gamma_yz}
\end{figure}

Figure \ref{fig_gamma_yz}(a) shows a representative effective index
distribution for $\hTM$ supermodes, taken at a slab thickness
$t_{\text{slab}}=190$ nm that is in the middle of the simulation
range. For both $y$- and $z$- dipole excitation, two supermodes,
labeled $\amode$ and $\bmode$, have the largest contribution to the
total collected PL. Representative electric-field profiles and
$\text{Re}\{n_\text{eff}\}$ for these main supermodes are shown in
Figs.~\ref{fig_mode_profiles}(d), ~\ref{fig_mode_profiles}(e) and
~\ref{fig_mode_profiles}(f). The type-$\amode$ supermode is well
confined, having the lowest $\text{Im}\{n_{\text{eff}}\}$, typically
an order of magnitude lower than its type-$\bmode$ counterpart and
all additional leaky modes. Leaky supermodes, pointed out in
Fig.~\ref{fig_gamma_yz}(a), provide a very small contribution to the
collected PL, typically at least two orders of magnitude lower than
the supermode with the largest contribution to $\eta_\text{PL}$. The
type-$\amode$ supermode is mostly concentrated in the fiber for
small $t_{\text{slab}}$, and in the slab at larger thicknesses, as
apparent in its fiber fraction plotted in
Fig.~\ref{fig_gamma_yz}(b). The type-$\bmode$ supermode resides
mostly in the fiber region for all slab thicknesses, although its
fiber-fraction increases with $t_{\text{slab}}$.

The individual supermode contributions $\eta_{\text{PL},m}$ to the
total PL collection for a $y$-dipole are plotted in
Fig.~\ref{fig_gamma_yz}(c). The type-$\amode$ supermode contribution
dominates at all thicknesses, while most leaky mode contributions
range below $10^{-3}$. Its modal emission ratio is maximized for
$t_{\text{slab}}\lessapprox$ 200 nm (Fig.~\ref{fig_gamma_yz}(d)),
for which a strong emission inhibition is observed
(Fig.~\ref{fig_total_rates}(a)). In this regime,
$\gamma_{\text{0,\amode}}$ is as high as 15~\% of the total emission
rate ($30~\%$ for emission in both $\pm z$ directions ). The
decreased type-$\amode$ contribution at larger $t_{\text{slab}}$ is
due to a reduction of both its fiber-fraction $f_{0,\amode}$ and
modal emission rate $\gamma_{0,\amode}$, as shown in
Figs.~\ref{fig_gamma_yz}(b) and ~\ref{fig_gamma_yz}(d),
respectively. Along with its increased fiber fraction
$f_{0,\bmode}$, this results in the type-$\bmode$ supermode power
approaching that of the type-$\amode$ supermode, in spite of its
significantly reduced $\gamma_{0,\bmode}$.

In comparison, the low $\eta_{\text{PL}}$ obtained for $z$-polarized
dipoles (Fig.~\ref{fig_efficiency}(c)) is a consequence of the small
mode emission rates (Figs.~\ref{fig_gamma_yz}(e) and
~\ref{fig_gamma_yz}(f)). In both $\hTM$ and $\hTE$ modes, the
$z$-electric field supermode components are dwarfed by the dominant,
$x$- or $y$-field components, resulting in small effective areas for
$z$-directed dipole moments. We note that the relatively poor
coupling of $z$-dipoles can prove polarization sensitivity which
can, for example, be used to determine the in-plane polarization of
an embedded semiconductor quantum dot.

As was the case for the $x$-dipole, $\eta_{\text{PL}}$ oscillates as
a function of $z$, with a period that is dominated by the beating of
these two main supermodes.  In addition, $\eta_{\text{PL}}$
decreases as a function of $z$, due to the decay of the various
supermodes.  However, this decay is significantly slower than that
found in the $x$-dipole case, as leaky modes play a far less
prominent role in PL collection here.  Over all slab thicknesses,
$\Rneff\gtrapprox1440$ for type-$\bmode$ supermodes, corresponding
to $N_{\delta=0.9}\lessapprox24$ free-space wavelengths for a 10~\%
amplitude decay. A coupling length shorter than $31\mum$ thus allows
for collection efficiencies at levels close to those plotted in
Figs.~\ref{fig_efficiency} (b) and ~\ref{fig_efficiency}(c). For the
type-$\amode$ supermode, $N_{\delta=0.9}\gtrapprox554$, so that a
collection level at least as high this supermode's individual
contributions (Figs.~\ref{fig_gamma_yz}(c) and
~\ref{fig_gamma_yz}(e)) may be achieved over coupling lengths of
several hundreds of microns.

\section{Discussion}
\label{section_discussion}

The results of Section \ref{section_analysis} indicate that
laterally bound or quasi-bound hybrid supermodes play an essential
role in luminescence collection in the single-emitter probing
technique described in this work (and schematically summarized in
Fig.~\ref{fig-sexy-scheme}). The overall PL collection efficiency
depends on four interdependent parameters for each supermode: total
($\Gamma$) and modal ($\Gamma_m$) spontaneous radiation rates,
fiber-mode fraction ($\f_m$), and the supermode decay rate
($\alpha_{z,m}$). High fiber-mode fractions may be achieved by
phase-matching the fiber and slab modes, however, a power shift
toward the fiber can result in a reduced field concentration at the
dipole position, and therefore smaller modal emission rates. A
compromise between $\Gamma_m$ and $\f_m$ determines the optimal
collection efficiency. The decay rate $\alpha_{z,m}$ determines the
length over which supermode collection is most efficient.

\begin{figure}[h]
\centerline{\includegraphics[scale=0.7,trim = 20 10 0
10]{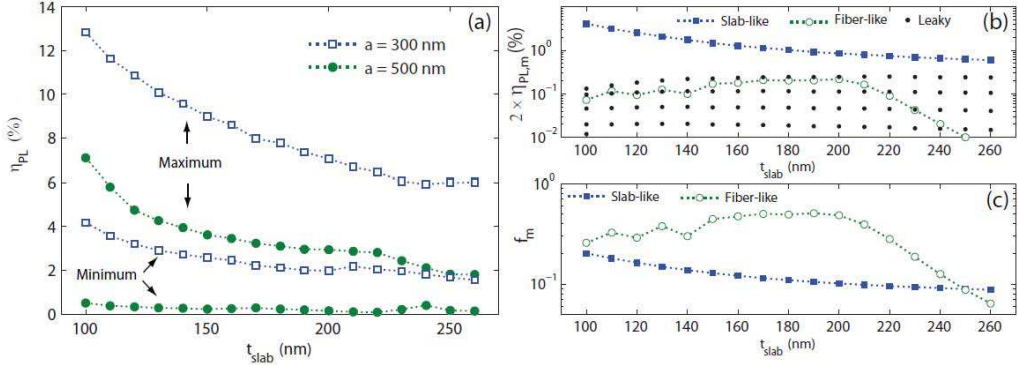}} \caption{ (a) Total PL collection
efficiencies (maximum and minimum values) for 300 nm and 500 nm
radius fibers and an $x$-oriented dipole moment at
$\lambda=1.3\mum$. (b) PL collection contributions from the
individual supermodes of the 300 nm radius fiber. The factor of 2 is
to account for collection from both fiber ends. Filled squares/empty
circles are for type $\amode$/$\bmode$ supermodes; black dots are
for leaky modes. (c) Fiber fractions for type-$\amode$ (filled
squares) and type-$\bmode$ (empty circles) supermodes of the 300-nm
radius fiber. } \label{fig-r300-eta-PL}
\end{figure}

A first method to obtain a substantially enhanced collection
efficiency is through a smaller diameter fiber. In
Fig.~\ref{fig-r300-eta-PL}(a), the maximum and minimum
$\eta_\text{PL}$ for an $x$-oriented dipole are plotted against slab
thickness, for a 300 $\nm$ radius fiber (the same procedure which
produced Fig.~\ref{fig_efficiency} was used, and the 500 nm radius
results are re-plotted here). The maximum PL collection is in
average 2.5 times that for a 500 $\nm$ fiber, while the minimum
efficiencies are increased in average by one order of magnitude, and
vary between 2$~\%$ and 4$~\%$. The collection enhancement is due to
the increased slab-like and leaky supermode PL contributions
(Fig.~\ref{fig-r300-eta-PL}(b)), which result from increased fiber
fractions (Fig.~\ref{fig-r300-eta-PL}(c)) that are enough to
compensate for a small decrease in the modal spontaneous emission
rates. In addition, leaky modes play a more important role than in
the $R=500\nm$ case, especially for $t_\text{slab}\gtrapprox180\nm$.
Among the leaky supermodes with $\eta_\text{PL,m}$ above 1~\% of the
slab-like supermode's, the one with the lowest individual
contribution has $\Rneff\lessapprox167$, or
$N_{\delta=0.9}\gtrapprox3$ wavelengths for 10~\% amplitude decay.
As such, the maximum collection efficiencies in
Fig.~\ref{fig-r300-eta-PL}(b) will occur for coupling lengths that
are within several microns. The slab-like supermode has
$N_{\delta=0.9}\gtrapprox382$, so collection at least as high as its
contribution in Fig.~\ref{fig-r300-eta-PL}(b) is achievable over
lengths shorter than $\approx$ 500$\mum$.

Our calculations also show good prospects for probing InAs quantum
dots in non-suspended slab configurations, where a
Al$_{0.7}$Ga$_{0.3}$As (refractive index $\approx 3.0$ at
$\lambda=1.3\mum$) substrate is present underneath the GaAs host
layer. For a slab thickness of 250 nm, we find, similarly to the
suspended membrane case, a slab-like supermode, several
slab-concentrated leaky supermodes, and a fiber-like supermode,
which in total yield a maximum collection efficiency of $\approx
1.6~\%$. The reduced collection with respect to the GaAs membrane
case is primarily due to a reduction in the supermode effective
areas, as fields extend over longer lengths inside the substrate. In
addition, the supermode decay lengths significantly change. The
fiber-like supermode, which provides the largest contribution to PL
collection, drops to 90~\% of its original amplitude in
approximately $2\mum$. Probing lengths of a few microns are thus
required for highly efficient PL collection. The slab-like and first
slab-concentrated leaky supermodes have $N_{\delta=0.9}\approx254$
and $N_{\delta=0.9}\approx24$, respectively. For coupling lengths as
long as $\approx31\mum$, the collection efficiency is at least
0.13~\%, the contribution of these two supermodes alone.

While the above cases and those studied in the rest of this paper
largely involve a tradeoff between $\Gamma_m$ and $\f_m$, situations
exist in which spontaneous emission enhancement and high
$\eta_{\text{PL}}$ can be reached through small effective area, high
fiber-mode fraction supermodes.  For example, this possibility
arises for the $\hTM$ modes if a $y$-oriented dipole is placed
between the fiber and slab. At $t_{\text{slab}}\approx165\nm$,
strong hybridization results in $y$-polarized supermodes with high
confinement in the air gap, similar to the slot waveguide modes
introduced in ~\cite{Almeida.OL.04}. The spontaneous emission rate
into this supermode alone reaches $\approx 1.7$ times the
homogeneous free-space rate, and the phase-matching between fiber
and slab leads to a fiber-mode fraction close to 50~\%. This
indicates promising possibilities for fluorescence collection from
dipoles located near the slab surface. Further investigation on
single emitter spectroscopy applications utilizing this
configuration is under way \cite{future_paper}.

For the case of $x$-oriented dipoles, the collection efficiency may
be effectively enhanced through stronger lateral confinement. In the
$\hTE$ mode case of Section \ref{section_analysis}, PL collection
efficiency is limited by the small fiber-mode fraction for the
slab-like supermode and the even smaller fiber-mode fraction (and
smaller spontaneous emission rates) for the slab-guided, leaky
supermodes.  To address these limitations, the slab may be converted
into a suspended channel waveguide, which has been fabricated in a
number of material systems, including recent experiments involving
semiconductor quantum dots \cite{koseki.APL}. This structure
provides guided supermodes with strong field confinement in the
channel, so that the supermodes carry most of the emitted power, and
may be designed to have high fiber-mode fractions, by phase-matching
fiber and channel modes. As an example, for a channel thickness
$t_{\text{channel}}=256\nm$ and width $W_{\text{channel}}=250\nm$,
two bound modes exist and carry $\approx 80~\%$ of the total
radiated power by an $x$-oriented dipole at the channel center. A
total PL collection efficiency in excess of $70~\%$ results.  A
detailed analysis of this system has been prepared
\cite{future_paper2}.

Finally, we consider the potential for resonant fluorescence
measurements within our fiber-based probing scheme.  As mentioned
previously, resonant spectroscopy offers many advantages in
comparison to non-resonant PL measurements, most notably improved
spectral resolution, avoidance of generation of decoherence-inducing
excess carriers and well-defined state preparation
\cite{gerardot_APL,vamivakas.nano.letters.7.2892,
Wrigge.Nature.Phys.4.60,gerhardt.prl.033601,srinivasan.pra.033839,Muller.prl}.
One of the main challenges in performing this type of spectroscopy
is effectively isolating the residual or scattered resonant
excitation beam from the resonance fluorescence signal, as they
occur at identical wavelengths (such separation is easily
accomplished in non-resonant PL through a grating spectrometer or
spectral filter). Recently, free-space optics collection of
resonance fluorescence from a quantum dot inside a planar cavity was
demonstrated \cite{Muller.prl}. In this work, separation of the
fluorescence signal from the excitation beam was accomplished
through an orthogonal excitation-collection scheme, where the sample
was excited in-plane through a cleaved optical fiber, while the
vertically emitted fluorescence was collected with a microscope
objective.

Through the setup of Fig.~\ref{fig-coupler}(a), isolation of a
resonant fluorescence signal can be achieved in a similar way. In a
resonance fluorescence experiment, the pump signal (indicated in
green in the figure) is tuned to the emitter transition. Illuminated
by the resonant pump, the emitter radiates in both forward and
backward directions of the fiber waveguide. The backward signal is
routed to an external detector via a directional coupler,
illustrated in Fig.~\ref{fig-coupler}(a), and is well-isolated from
the pump signal (a fiber optic circulator could alternatively be
used). Given the high fiber-collection efficiencies predicted in
Fig.~\ref{fig_efficiency} (the collection in the backward channel
will be 0.5$\eta_{\text{PL}}$), this configuration may provide
substantially superior signal-to-noise ratios in comparison to
schemes that rely on free-space fluorescence collection.
Alternately, we note that one might envision exciting the sample
with a resonant free-space beam normal to the sample, while
collecting the fluorescence through both the forward and backward
channels of the fiber.  This would create an orthogonal
excitation-collection scheme similar to that of ~\cite{Muller.prl},
with an overall efficiency given by $\eta_{\text{PL}}$ as plotted in
Fig.~\ref{fig_efficiency}. Such a configuration may be preferable in
situations where reflections in the fiber path between source and
emitter can not be avoided.

\section{Summary and Conclusions}
\label{section_conclusion}

A technique based on the use of an optical fiber taper waveguide for
collecting emission from single emitters embedded in thin dielectric
membranes was analyzed with numerical simulations. The probing
configuration was modeled as a composite waveguide formed by a
micron-scale silica fiber on top of a dielectric slab.
Finite-difference time-domain simulations were used to estimate the
efficiency with which emission from in-plane and vertically-oriented
dipoles can be collected into the fundamental mode of the fiber. For
in-plane dipoles, collection efficiencies superior to those
achievable with a high numerical aperture objective are predicted,
by as much as an order of magnitude. A maximum collection efficiency
of $\approx 7\%$ may be obtained, with insignificant radiation
suppression due to the slab. For vertically oriented dipoles,
collection efficiency superior to free-space collection by at least
an order of magnitude is predicted. A maximum collection above
$35~\%$ is obtained, albeit with strong slab-induced radiation
suppression (Purcell factor of $<0.05$). It is important to note
that the presence of the fiber does not significantly affect the
radiative rate of the slab-embedded dipole, so the suppression
equally affects free-space collection.

Finite element simulations were used to understand the FDTD results
through contributions from guided and leaky supermodes of the
composite waveguide formed by the fiber and slab. Our analysis
determined the essential role of laterally bound or quasi-bound
hybrid supermodes in luminescence collection, as well as the
relevant parameters for increased efficiency: total ($\Gamma$) and
modal ($\Gamma_m$) spontaneous radiation rates, fiber-mode fraction
($\f_m$), and the supermode decay rate ($\alpha_{z,m}$). In short,
the ratio $\Gamma_m/\Gamma$ (or, the supermode $\beta$-factor) and
$\f_m$ must be maximized, while $\alpha_{z,m}$ must be minimized for
higher collection levels. Potential methods to further increase the
collection efficiency, based on supermode analysis, have been
introduced. Furthermore, probing of single quantum dots in
unprocessed (non-undercut) samples is predicted to yield collection
levels on the same order as obtained with free-space collection.
Finally, we have described how the probing method is amenable to
both non-resonant photoluminescence and resonant fluorescence
measurements.

In addition to efficient collection, optical fiber taper waveguides
offer a number of benefits in the study of single solid-state
emitters.  These devices have standard single mode fiber input and
output regions, and are therefore easily integrated with the
technology of low-loss fiber optics. As discussed elsewhere
\cite{srinivasan.pra.033839}, low-temperature measurement setups
incorporating fiber taper waveguides and other fiber optic
components have been used in cavity QED experiments in which precise
knowledge of optical losses and power levels are a requirement.
Furthermore, we note that as a movable probe, fiber taper waveguides
have been used to interrogate two-dimensional arrays of devices on a
chip \cite{michael.OE}.  The results of this article indicate
that these advantageous features, previously demonstrated within the
context of microcavity spectroscopy, can be utilized in direct
single emitter spectroscopy.

\renewcommand{\theequation}{A-\arabic{equation}} 
\setcounter{equation}{0}  
\section*{Appendix A - Classical point dipole radiation in the presence of a waveguide}
In this section, we demonstrate the validity of
Eq.~(\ref{eq_spont_rate_frac}), which states that the ratio of the
spontaneous emission rate in bulk to that in a waveguide is equal to
the ratio of the total classical dipole radiated power in bulk to
that in a waveguide.  We consider a point dipole source
$\mathbf{J}(\mathbf{r},t)=-i\omega\mathbf{p} \delta (\mathbf{r})
e^{-i\omega_0t}$ radiating in the presence of an arbitrary, lossless
dielectric waveguide extending along the $z$ direction. The
electromagnetic field may be expanded in terms of the bound and
radiative modes of the waveguides as \cite{snyder.love.waveguides}:
\begin{eqnarray}
\left\{ \begin{array}{c} \mathbf{E} \\ \mathbf{H}
\end{array} \right\} = \sum_{j}\left[
a_j(z) e^{i\beta_jz}+a_{-j}(z)e^{-i\beta_jz}\right]
  \left\{ \begin{array}{c} \mathbf{e}_j \\ \mathbf{h}_j \end{array}\right\}+\nonumber\\
\sum_{j}\int d\beta\,\left[ a_j(z,\beta) e^{i\beta
z}+a_{-j}(z,\beta)e^{-i\beta z} \right]\left\{
\begin{array}{c}
\mathbf{e}_j(\beta) \\ \mathbf{h}_j(\beta)
\end{array}\right\}\label{eq_E_field_exp}
\end{eqnarray}
with amplitude coefficients
\begin{equation}
a_{\pm j}(z)= \begin{cases} \ 0 &\text{for $z \gtrless 0$}\\
i\frac{\omega_0\mathbf{p}\cdot\mathbf{e}^*_j}{4N_j} &\text{for
$z\gtrless 0$}
\end{cases}\quad\text{and}\quad
a_{\pm j}(z,\beta)= \begin{cases} \ 0 &\text{for $z \gtrless 0$}\\
i\frac{\omega_0\mathbf{p}\cdot\mathbf{e}^*_j(\beta)}{4N_j(\beta)}
&\text{for $z\gtrless 0$}
\end{cases}.
\end{equation}
The modes are such that the following orthogonality relations are
valid:
\begin{equation}
\left|\iint\limits_S\left(\mathbf{e}_j\times\mathbf{h}_k^*
\right)\cdot \hat{\mathbf{z}}\right|=N_j\,\delta_{j,k},
\end{equation}
\begin{equation}
\left|\iint\limits_S\left[\mathbf{e}_j(\beta)\times\mathbf{h}_k^*(\beta')
\right]\cdot
\hat{\mathbf{z}}\right|=N_j(\beta)\,\delta(\beta-\beta')\delta_{j,k}.
\end{equation}
The total power flowing through the waveguide at a position $z>0$ is
obtained by integrating
$1/2\left(\mathbf{E}\times\mathbf{H}^*\right)\cdot\mathbf{z}$ over
the cross-section:
\begin{eqnarray}
\mathbf{S}_z= \frac{|\mathbf{p}|^2\omega_0^2}{2}\sum_{j}\left\{
\frac{|\hat{\mathbf{p}}\cdot\mathbf{e}_j|^2}{N_j}+ \int
d\beta\,\frac{|\hat{\mathbf{p}}\cdot\mathbf{e}_j(\beta)|^2}{N_j(\beta)}\right\}_{\mathbf{r}=\mathbf{0}}
\label{eq_Sz_waveguide}
\end{eqnarray}
We next normalize the guided and radiative mode fields as in
Eqs.~(\ref{eq_guided_mode_norm}) and (\ref{eq_radiative_mode_norm}).
The guided mode propagation constants $\beta_j$ may be expressed as
\cite{snyder.love.waveguides}
\begin{equation}
\beta_j = \omega_0\mu_0\frac{\iint\limits_S dS\,\epsilon(\mathbf{r})\left(
\mathbf{e}_j\times\mathbf{h}_j^*\right)\cdot\hat{\mathbf{z}}
}{\iint\limits_S dS\,\epsilon(\mathbf{r})|\mathbf{e}_j|^2}=\iint\limits_S
dS\,\epsilon(\mathbf{r})\left(
\mathbf{e}_j\times\mathbf{h}_j^*\right)\cdot\hat{\mathbf{z}}
\label{eq_beta_j}
\end{equation}
Considering the waveguide material to be non-dispersive, the modal
group velocity may be written as \cite{snyder.love.waveguides}
\begin{equation}
v_{g,j} = \left( \frac{d\beta_j}{d\omega_0} \right)^{-1} =
c\beta_j\left(\frac{c}{\omega_0}\right)\frac{N_j}{\iint\limits_S
dS\,\epsilon(\mathbf{r})\left( \mathbf{e}_j\times\mathbf{h}_j^*
\right)\cdot \hat{\mathbf{z}}}. \label{eq_vg_j}
\end{equation}
Using Eqs. (\ref{eq_vg_j}) and (\ref{eq_beta_j}) and
Eq.~(\ref{eq_Sz_waveguide}), the power through the waveguide
cross-section, normalized to the homogeneous dipole power,
Eq.~(\ref{eq_hom_radiated_power}) becomes:
\begin{eqnarray}
\mathbf{S}_z= \sum_{j}\left\{ \frac{6\pi c^3\beta_j'}{n\omega_0^2}
|\hat{\mathbf{p}}\cdot\mathbf{e}_j|^2+\int d\beta\, \frac{6\pi
c}{\mu_0n\omega_0^2}
\frac{|\hat{\mathbf{p}}\cdot\mathbf{e}_j(\beta)|^2}
{N_j(\beta)}\right\}_{\mathbf{r}=\mathbf{0}}.
\label{eq_Sz_waveguide}
\end{eqnarray}
The same expression is obtained by normalizing the spontaneous
emission rate $\Gamma$ in Eq.~(\ref{eq_Gamma}) by the spontaneous
emission rate of a dipole in a homogeneous dielectric medium, given
by:

\begin{equation}
\Gamma_{\text{Hom.}} =
\frac{\omega_0^3|\mathbf{d}_{eg}|^2n}{3\pi\hbar\epsilon_0c^3}.
\label{eq_hom_spont_rate}
\end{equation}

\begin{figure}[h]
\centerline{\includegraphics[scale=0.6,trim = 0 10 10
25]{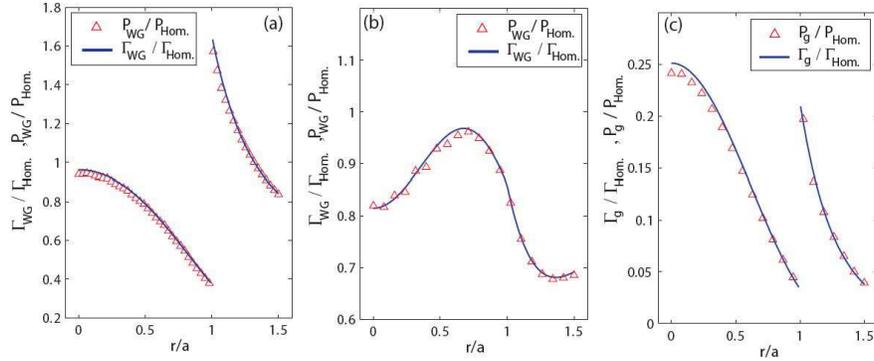}} \caption{(a) Emitted power and modal spontaneous
emission rate at $\lambda=1.56 \mum$, normalized to the
corresponding quantities in a homogeneous medium, for a (a) radially
and (b) longitudinally oriented dipole at position $r$ in the fiber
described in the text. (c) Emitted guided mode power, $P_g$, and
spontaneous emission rate, $\Gamma_g$, normalized to the total
emitted power and emission rates, at $\lambda=1.56\mum$ for a
radially oriented dipole at position $r$ in the fiber described in
the text.} \label{fig_compare_fdtd_qm}
\end{figure}

To verify Eq.~(\ref{eq_spont_rate_frac}), we used FDTD to simulate a
point dipole radiating at $\lambda=1.56\mum$ near an optical fiber
of index $n=1.45$ and radius $a=500\nm$, surrounded by air. The
radiated power ratio $P_{\text{WG}}/P_{\text{Hom.}}$ is compared to
the spontaneous emission rate ratio
$\Gamma_{\text{WG}}/\Gamma_{\text{Hom.}}$, calculated with the
semi-analytical expressions of ~\cite{LeKien.pra.72.032509}.
Figures~\ref{fig_compare_fdtd_qm}(a) and
~\ref{fig_compare_fdtd_qm}(b) show the two ratios for radially and
longitudinally oriented dipoles at position $r$ along the radial
direction.

The relation between the classical guided fiber mode power
$P_g/P_{\text{WG}}$ and spontaneous emission rate
$\Gamma_g/\Gamma_{\text{WG}}$ ratios was also verified. The
steady-state fields recorded at one of the waveguide edges were used
in the overlap integral of Eq.~(\ref{eq_overlap}), together with the
fundamental fiber mode fields, to yield $P_g$. Figure
\ref{fig_compare_fdtd_qm}(c) compares the ratios for the
radially-oriented dipole as a function of its position $r$ in the
radial direction.

\renewcommand{\theequation}{B-\arabic{equation}} 
\setcounter{equation}{0}  
\section*{Appendix B - Heisenberg equations}
To describe the spontaneous emission of a two-level emitter in the
vicinity of a dielectric waveguide, we obtain the Heisenberg
equations for the emitter interacting with a vacuum field reservoir
that is described in terms of the guided and radiative waveguide
modes propagating in the $z$-direction.

Through the quantization procedure of ~\cite{gardiner.pra.31.3761},
the positive-frequency electric field operators are defined as
$\mathbf{E}^{(+)}=\mathbf{E}_{\text{guided}}^{(+)}+\mathbf{E}_{\text{radiative}}^{(+)}$,
where
\begin{equation}
\mathbf{E}_{\text{guided}}^{(+)}= i\sum_{f,m}\int_0^\infty
d\omega\,\sqrt{\frac{\hbar\omega\beta_\mu'}{4\pi\epsilon_0}}
a_\mu\mathbf{e}_\mu e^{-i(\omega t-f\beta_\mu z)} \label{eq_Eg}
\end{equation}
\begin{equation}
\mathbf{E}_{\text{radiative}}^{(+)}= i\sum_{n}\int_0^\infty
d\omega\,\int d\beta\sqrt{\frac{\hbar\omega}{4\pi N_\nu}}
a_\nu\mathbf{e}_\nu e^{-i(\omega t-\beta z)} \label{eq_Er}
\end{equation}
Here, as in Section \ref{section_analysis}, $\mu=(f,m,\omega)$
labels a guided mode of order $m$ traveling in the direction $f\cdot
z$, $f=\pm 1$, with propagation constant $\beta_{\mu}$, and inverse
group velocity $\beta_\mu'=\frac{d\beta_\mu}{d\omega}=v_g^{-1}$.
$\mathbf{e}_\mu$ is the mode field profile and $a_{\mu}$ is the
corresponding photon annihilation operator. In (\ref{eq_Er}),
$\nu=(n,\beta,\omega)$ labels a radiative mode of order $n$ and
propagation constant $\beta$, with $\mathbf{e}_\nu$ the
corresponding field profile and $a_{\nu}$ the photon annihilation
operator. The $\beta$-integral limits are such that evanescent
radiation modes are excluded (e.g., $|\beta|<\omega/c$ if the
waveguide is embedded in air) \cite{sondegaard.PhysRevA.64.033812}.
Also, $N_\nu$ is such that
\begin{equation}
\iint\limits_S dS\,\left(
\mathbf{e}_\nu\times\mathbf{h}_{\nu'}^*\right)\cdot\hat{\mathbf{z}}=N_\nu\delta(\beta-\beta')
.
\end{equation}
The guided and radiative field profiles are normalized as in Section
\ref{section_analysis}, such that $\left[
a_\mu,a_{\mu'}\right]=\delta(\omega-\omega')\delta_{ff'}\delta_{mm'}$
and $\left[
a_\nu,a_{\nu'}\right]=\delta(\omega-\omega')\delta(\beta-\beta')\delta_{nn'}$.

Next, consider a two-level atom (emitter) in the dipole
approximation, interacting with a photon reservoir. The atom has a
dipole moment $\mathbf{d}_{eg}$ for the transition with energy
$\hbar\omega_0$ between the ground $\ket{g}$ and excited $\ket{e}$
states. We define the atomic transition operators
$\sigma_-=\ket{g}\bra{e}$, $\sigma_+=\ket{e}\bra{g}$,
$\sigma_z=\ket{e}\bra{e}-\ket{g}\bra{g}$, such that the atomic
Hamiltonian is $H_A = \hbar\omega_0\sigma_z/2$. The total
Hamiltonian for the interacting atom and reservoir is
\begin{equation}
 H = H_A + H_R + H_{I_R}.
\end{equation}
The interaction Hamiltonian for electric dipole coupling is given by
$H_{I_R}=-\mathbf{d}_{eg}\cdot\mathbf{E}$, which, in the
rotating-wave approximation, may be written as
\cite{LeKien.pra.72.032509}:
\begin{eqnarray}
\lefteqn{H_{I_R} = i\hbar \left\{ \sum_{f,m}\int_0^\infty d\omega\,
\left[ G_\mu^*a_\mu^\dagger\sigma_-e^{i(\omega-\omega_0)t}
-G_\mu\sigma_+a_\mu e^{-i(\omega-\omega_0)t} \right] \right\} }\nonumber\\
&&+i\hbar\left\{ \sum_{n}\int_0^\infty d\omega\,\int d\beta\, \left[
G_\nu^*a_\nu^\dagger\sigma_-e^{i(\omega-\omega_0)t}
-G_\nu\sigma_+a_\nu e^{-i(\omega-\omega_0)t} \right] \right\},
\label{eq_Hint}
\end{eqnarray}
with
\begin{eqnarray}
G_\mu = \sqrt{\frac{\omega\beta_\mu'}{4\pi\epsilon_0\hbar}}\left
(\mathbf{d}_{eg}\cdot\mathbf{e}_\mu\right)\label{eq_G_mu}
\qquad\text{and}\qquad G_\nu = \sqrt{\frac{\omega}{4\pi
N_\nu\hbar}}\left (\mathbf{d}_{eg}\cdot\mathbf{e}_\nu\right).
\end{eqnarray}
The formal solutions of the field operators are
\cite{cohen.tannoudji.atom.photon}:
\begin{equation}
a_{\mu,\nu}=a_{\mu,\nu}(t_0)+G_{\mu,\nu}^*\int_{t_0}^t
dt'\,\sigma_-e^{i(\omega -\omega_0)t'}.\label{eq_a_mu}
\end{equation}
Under the Markoff approximations $\sigma_-(t-\tau)\to\sigma_-(t)$
and $G_{\mu,\nu}(\omega) \approx G_{\mu,\nu}(\omega_0)$, the
following Heisenberg equations for the atomic operators may be
obtained:
\begin{equation}
\frac{d\tilde{\sigma}_-}{dt}=-i\omega_0\tilde{\sigma}_--\frac{\Gamma}{2}\tilde{\sigma}_-+\sigma_z\left(\sum_{f,m}
\sqrt{\Gamma_{\mu}}g_{\mu, \text{in}}(t)+\sum_{n} \int d\beta\,
\sqrt{\Gamma_{\nu}} r_{\nu, \text{in}}(t)\right),
\label{eq_heisenberg_sigma_minus}
\end{equation}
\begin{eqnarray}
\frac{d\sigma_z}{dt}& = &-\Gamma(1+\sigma_z) \nonumber  -
2\tilde{\sigma}_+\left(\sum_{f,m} \sqrt{\Gamma_{\mu}}g_{\mu,
\text{in}}(t)+\sum_{n}
\sqrt{\Gamma_{\nu}}r_{\nu, \text{in}}(t)\right) \nonumber \\
& - &2\tilde{\sigma}_-\left(\sum_{f,m} \sqrt{\Gamma_{\mu}}^*g_{\mu,
\text{in}}^\dagger(t)+\sum_{n} \int d\beta\,
\sqrt{\Gamma_{\nu}}^*r_{\nu, \text{in}}^\dagger(t)\right),
\label{eq_heisenberg_sigmaz}
\end{eqnarray}
with $\sigma_-=\tilde{\sigma}_-e^{i\omega_0t}$,
\begin{equation}
\Gamma = \left(\sum_{f,m}\Gamma_\mu+\sum_{n}\int
d\beta\,\Gamma_\nu\right),\label{eq_Gamma}
\end{equation}
\begin{equation}
\sqrt{\Gamma_\mu} = \sqrt{2\pi}G_\mu(\omega_0)
\qquad\text{and}\qquad \sqrt{\Gamma_\nu} =
\sqrt{2\pi}G_\nu(\omega_0).\label{eq_Gamma_nu}
\end{equation}
To arrive at these results, the following input field operators have
been defined \cite{gardiner.pra.31.3761}:
\begin{equation} \left\{ \begin{array}{c} g_{\mu, \text{in}}(t) \\ r_{\nu, \text{in}}(t) \end{array} \right\}=
\frac{1}{\sqrt{2\pi}}\int_{-\infty}^{+\infty}d\omega \,\left\{
\begin{array}{c} \tilde{a}_{\mu}(t_0) \\ \tilde{a}_{\nu}(t_0) \end{array} \right\} e^{-i\omega(t-t_0)}
\end{equation}
Here, $a_{\mu,\nu}(t_0)=\tilde{a}_{\mu,\nu}(t_0)e^{i\omega t_0}$.

\section*{Acknowledgement}
This work has been supported in part by the NIST-CNST/UMD-NanoCenter
Cooperative Agreement.


\begin{thebibliography}{}

\bibitem{gerardot_APL}
B.~Gerardot, S.~Seidl, P.~Dalgarno, R.~Warburton, M.~Kroner,
K.~Karrai,
  A.~Badolato, and P.~Petroff, ``Contrast in transmission spectroscopy
  of a single quantum dot,'' Appl. Phys. Lett. {\bf 90}, 221\,106 (2007).

\bibitem{vamivakas.nano.letters.7.2892}
A.~N. Vamivakas, M.~Atature, J.~Dreiser, S.~T. Yilmaz, A.~Badolato,
A.~K. Swan,
  B.~B. Goldberg, A.~Imamoglu, and M.~S. Unlu, ``Strong Extinction of a
  Far-Field Laser Beam by a Single Quantum Dot,'' Nano Lett. {\bf 7},
  2892--2896 (2007).

\bibitem{Wrigge.Nature.Phys.4.60}
G.~Wrigge, I.~Gerhardt, J.~Hwang, G.~Zumofen, and V.~Sandoghdar,
  ``Efficient coupling of photons to a single molecule and the
  observation of its resonance fluorescence,'' Nat. Phys. {\bf 4}, 60--66
  (2008).

\bibitem{gerhardt.prl.033601}
I.~Gerhardt, G.~Wrigge, P.~Bushev, G.~Zumofen, M.~Agio, R.~Pfab, and
  V.~Sandoghdar, ``Strong Extinction of a Laser Beam by a Single
  Molecule,'' Phys. Rev. Lett. {\bf 98}, 033\,601 (2007).

\bibitem{aoki.nat}
T.~Aoki, B.~Dayan, E.~Wilcut, W.~P. Bowen, A.~S. Parkins, H.~J.
Kimble, T.~J.
  Kippenberg, and K.~J. Vahala, ``Observation of Strong Coupling between
  One Atom and a Monolithic Microresonator,'' Nature {\bf 443}, 671--674 (2006).

\bibitem{srinivasan.nat}
K.~Srinivasan and O.~Painter, ``Linear and nonlinear optical
  spectroscopy of a strongly coupled microdisk-quantum dot system,'' Nature {\bf
  450}, 862--865 (2007).

\bibitem{srinivasan.pra.033839}
K.~Srinivasan, C.~P. Michael, R.~Perahia, and O.~Painter,
  ``{Investigations of a coherently driven semiconductor optical cavity
  QED system},'' Phys. Rev. A {\bf 78}, 033\,839 (2008).

\bibitem{LeKien.pra.72.032509}
F.~Le~Kien, S.~Dutta~Gupta, V.~I. Balykin, and K.~Hakuta,
``Spontaneous
  emission of a cesium atom near a nanofiber: Efficient coupling of light to
  guided modes,'' Phys. Rev. A {\bf 72}, 032\,509 (2005).

\bibitem{klimov.PhysRevA.69.013812}
V.~V. Klimov and M.~Ducloy, ``Spontaneous emission rate of an
excited
  atom placed near a nanofiber,'' Phys. Rev. A {\bf 69}, 013\,812 (2004).

\bibitem{Nayak_OE}
K.~Nayak, P.~Melentiev, M.~Morinaga, F.~Le~Kien, V.~Balykin, and
K.~Hakuta,
  ``Optical nanofiber as an efficient tool for manipulating and probing
  atomic fluorescences,'' Opt. Express {\bf 15}, 5431--5438 (2007).

\bibitem{srinivasan:091102}
K.~Srinivasan, O.~Painter, A.~Stintz, and S.~Krishna, ``Single
quantum
  dot spectroscopy using a fiber taper waveguide near-field optic,'' Appl. Phys.
  Lett. {\bf 91}, 091\,102 (2007).

\bibitem{Muller.prl}
A.~Muller, E.~B. Flagg, P.~Bianucci, X.~Wang, D.~G. Deppe, W.~Ma,
J.~Zhang,
  G.~J. Salamo, M.~Xiao, and C.~K. Shih, ``Resonance Fluorescence from a
  Coherently Driven Semiconductor Quantum Dot in a Cavity,'' Phys. Rev. Lett.
  {\bf 99}, 187\,402 (2007).

\bibitem{snyder.love.waveguides}
A.~W. Snyder and J.~D. Love, {\em Optical Waveguide Theory\/}
(Chapman and
  Hall, New York, 1983).

\bibitem{wang.APL.3423}
C.~F. Wang, A.~Badolato, I.~Wilson-Rae, P.~M. Petroff, E.~Hu,
J.~Urayama, and
  A.~Imamoglu, ``Optical properties of single InAs quantum dots in close
  proximity to surfaces,'' Appl. Phys. Lett. {\bf 85}, 3423--3425 (2004).

\bibitem{ref:Huang3}
W.-P. Huang, ``{Coupled-mode theory for optical waveguides: and
  overview},'' J. Opt. Soc. Am. A {\bf 11}, 963--983 (1994).

\bibitem{lumerical}
Lumerical FDTD Solutions. Specific software packages are identified
in this
  paper to foster understanding. Such identification does not imply
  recommendation or endorsement by NIST, nor does it imply that the software
  identified is necessarily the best available for the purpose.

\bibitem{Jackson.dipole.power}
J.~D. Jackson, {\em Classical Electrodynamics\/} (Wiley, New York,
1999), 3rd
  edn.

\bibitem{Benisty.JOSAA.98}
H.~Benisty, R.~Stanley, and M.~Mayer, ``Method of source terms for
  dipole emission modification in modes of arbitrary planar structures,'' J.
  Opt. Soc. Am. A {\bf 15}, 1192--1201 (1998).

\bibitem{Rigneault.PhysRevA.54.2356}
H.~Rigneault and S.~Monneret, ``Modal analysis of spontaneous
emission
  in a planar microcavity,'' Phys. Rev. A {\bf 54}, 2356--2368 (1996).

\bibitem{Urbach.PhysRevA.57.3913}
H.~P. Urbach and G.~L. J.~A. Rikken, ``Spontaneous emission from a
  dielectric slab,'' Phys. Rev. A {\bf 57}, 3913--3930 (1998).

\bibitem{michael.OE}
C.~P. Michael, M.~Borselli, T.~J. Johnson, and O.~Painter, ``An
optical
  fiber taper probe for wafer-scale microphotonic device characterization,''
  Opt. Express {\bf 15}, 4745--4752 (2007).

\bibitem{Jayavel}
P.~Jayavel, H.~Tanaka, T.~Kita, O.~Wada, H.~Ebe, M.~Sugawara,
J.~Tatebayashi,
  Y.~Arakawa, Y.~Nakat, and T.~Akiyama, ``{Control of optical
  polarization anisotropy in edge emitting luminescence of InAs/GaAs
  self-assemble quantum dots},'' Appl. Phys. Lett. {\bf 84} (2004).

\bibitem{sondegaard.PhysRevA.64.033812}
T.~S\o{}ndergaard and B.~Tromborg, ``General theory for spontaneous
  emission in active dielectric microstructures: Example of a fiber amplifier,''
  Phys. Rev. A {\bf 64}, 033\,812 (2001).

\bibitem{Xu:99}
Y.~Xu, J.~S. Vu\v{c}kovi\'{c}, R.~K. Lee, O.~J. Painter, A.~Scherer,
and
  A.~Yariv, ``Finite-difference time-domain calculation of spontaneous
  emission lifetime in a microcavity,'' J. Opt. Soc. Am. B {\bf 16}, 465--474
  (1999).

\bibitem{comsol}
Comsol Multiphysics. Specific software packages are identified in
this paper to
  foster understanding. Such identification does not imply recommendation or
  endorsement by NIST, nor does it imply that the software identified is
  necessarily the best available for the purpose.

\bibitem{Almeida.OL.04}
V.~R. Almeida, Q.~Xu, C.~A. Barrios, and M.~Lipson, ``Guiding and
  confining light in void nanostructure,'' Opt. Lett. {\bf 29}, 1209--1211
  (2004).

\bibitem{future_paper}
M.~Davan\c{c}o and K.~Srinivasan, ``Optical fiber taper waveguides
for
  highly efficient spectroscopy of single emitters deposited on a dielectric
  slab,''  Manuscript in preparation (2009).

\bibitem{koseki.APL}
S.~Koseki, B.~Zhang, K.~D. Greve, and Y.~Yamamoto, ``Monolithic
  integration of quantum dot containing microdisk microcavities coupled to
  air-suspended waveguides,'' Appl. Phys. Lett. {\bf 94}, 051\,110 (2009).

\bibitem{future_paper2}
M.~Davan\c{c}o and K.~Srinivasan, ``Fiber-coupled semiconductor
  waveguides as an efficient optical interface to a single quantum dipole,''
   preprint: {\it{arxiv.org/abs/0905.2994}} (2009).

\bibitem{gardiner.pra.31.3761}
C.~W. Gardiner and M.~J. Collett, ``Input and output in damped
quantum
  systems: Quantum stochastic differential equations and the master equation,''
  Phys. Rev. A {\bf 31}, 3761--3774 (1985).

\bibitem{cohen.tannoudji.atom.photon}
C.~Cohen-Tannoudji, J.~Dupont-Roc, and G.~Grynberg, {\em Atom-Photon
  Interactions: Basic Processes and Applications\/} (Wiley Interscience, New
  York, 1998).

\end{thebibliography}
\end{document}